\documentclass[letterpaper,10pt,times,mathptm,psfig,conference]{IEEEtran}
\IEEEoverridecommandlockouts
\usepackage{cite}
\usepackage{amsmath,amssymb,amsfonts}
\usepackage{algorithmic}
\usepackage[ruled,linesnumbered]{algorithm2e}
\usepackage[normalem]{ulem}
\usepackage{amssymb}
\usepackage{amsmath}
\usepackage{graphicx}
\usepackage{textcomp}
\usepackage{xcolor}
\usepackage{epsfig, subfigure, amsmath, amssymb, wrapfig}
\def\BibTeX{{\rm B\kern-.05em{\sc i\kern-.025em b}\kern-.08em
    T\kern-.1667em\lower.7ex\hbox{E}\kern-.125emX}}
    
\newcommand{\name}{{\textsc{\small{CURe}}}\xspace}
\newcommand{\namet}{{\textsc{CURe}}\xspace}

\newcommand{\done}[1]{}

\newcommand{\alvi}[1]{}

\newcommand{\rahman}[1]{}

\author{
Alvi Ataur Khalil\IEEEauthorrefmark{1}, Mohamed Y. Selim\IEEEauthorrefmark{2}, Mohammad Ashiqur Rahman\IEEEauthorrefmark{1}\\
\IEEEauthorrefmark{1}Analytics for Cyber Defense (ACyD) Lab, Florida International University, USA\\
\IEEEauthorrefmark{2}Department of Electrical and Computer Engineering, Iowa State University, USA\\
\IEEEauthorrefmark{1}\{akhal042, marahman\}@fiu.edu, \IEEEauthorrefmark{2}myoussef@iastate.edu
}

\begin{document}
\title{\namet{}: Enabling RF Energy Harvesting using Cell-Free Massive MIMO UAVs Assisted by RIS}



\maketitle
\thispagestyle{empty}
\pagestyle{empty}

\begin{abstract}
The ever-evolving internet of things (IoT) has led to the growth of numerous wireless sensors, communicating through the internet infrastructure. When designing a network using these sensors, one critical aspect is the longevity and self-sustainability of these devices. For extending the lifetime of these sensors, radio frequency energy harvesting (RFEH) technology has proved to be promising. In this paper, we propose \name{}, a novel framework for RFEH that effectively combines the benefits of cell-free massive MIMO (CFmMIMO), unmanned aerial vehicles (UAVs), and reconfigurable intelligent surfaces (RISs) to provide seamless energy harvesting to IoT devices. We consider UAV as an access point (AP) in the CFmMIMO framework. To enhance the signal strength of the RFEH and information transfer, we leverage RISs owing to their passive reflection capability. Based on an extensive simulation, we validate our framework's performance by comparing the max-min fairness (MMF) algorithm for the amount of harvested energy.

\end{abstract}


\section{Introduction}
\label{sec:introduction}


At present, it is estimated that more than 50 billion internet of things (IoT) devices are connected to the internet~\cite{kavyashree2018survey, balaji2019iot, haque2021novel}, including diverse domains such as home and living environment, communication and connectivity, agricultural, healthcare, and medical,  transportation and logistics, etc~\cite{alzubi2019survey, kotha2018iot, khalil2021efficient, khalil2021literature}. 
These IoT devices employ a variety of methodologies for connecting and sharing data, and most of these methods are wireless by nature~\cite{sikimic2020overview}. Also, the major portion of the IoT ecosystem is transportable, owing to their applications. So, the charging of these IoT nodes is a crucial factor for the continuous operability of these devices. Moreover, many applications require tiny wireless IoT nodes to be deployed in places that are inaccessible and, there is no permanent power supply either~\cite{zorbas2018computing}. Although the rechargeable batteries can work as a conventional power source, they can only last for a specific time interval. Afterward, these batteries will also require charging or even replacement, which may incur higher costs. To resolve the problem of stationary charging, a lot of attention had been put to the research works related to finding ways of delivering power wirelessly. The great visionary Nikola Tesla first proposed to transmit energy into free space and convert the energy into usable direct current power~\cite{huang2019wireless}. Later, that foresight led to the development of state-of-the-art power supply techniques like Wireless Power Transfer (WPT) and Energy Harvesting (EH).

The cutting edge of the wireless power delivery is the harvesting of energy from radio frequency (RF) signals, providing seamless power to the IoT nodes indoor and outdoor, regardless of whether they are stationary or mobile~\cite{ren2018rf}. However, the conversion efficiency of RF energy harvesting (RFEH) is still low, particularly in long-distance transmissions. For dealing with the distant problem arising from the deployment of access points (APs), many recent works consider adopting a cell-free network system with a central processing unit and fronthaul connection so that there will always be some APs close to each of the IoT nodes. Among them, one group of energy harvesting mechanisms consider the maximization of total RFEH throughput along with the AP selection under transmission power constraints for each AP (e.g.,\cite{tran2018green}). However, they assume perfect channel state information (CSI) without considering the uplink communications. Another group of works minimize the total transmitted energy for wirelessly-powered cell-free IoT by considering a linear energy harvesting model with only Rayleigh fading (e.g.,~\cite{wang2020wirelessly}). Instead, adopting a Rician fading channel model with random phase shifts to each of the antennas of the APs would have been more general like the real environment. 

As the massive multiple-input multiple-output (mMIMO) \done{I recommend using mMIMO then change it accordingly in the paper} technique has the ability to focus the received signal power with very narrow beams~\cite{marzetta2010noncooperative, zheng2015survey}, it has been considered in wireless energy and information transfer systems~\cite{yang2015throughput, zhao2015downlink}. The practicality of mMIMO's wireless energy transfer capability for sensor networks energy harvesting was analyzed by Kashyap et al.~\cite{kashyap2016feasibility}. Although the mMIMO technology provided better efficiency in data rates compared to the previous technique, there is the issue of inter-cell interference. Consequently, it will fail to provide the desired level of energy transfer efficiency to the IoT devices. To serve all devices in a coherent manner, an improved network infrastructure is considered by Ngo et al.~\cite{ngo2017cell} with cell-free mMIMOs (CFmMIMO), for APs distributed over the coverage area. Leveraging this infrastructure, we propose \name{}, a \textbf{C}FmMIMO-mount \textbf{U}nmanned aerial vehicle (UAV) assisted by \textbf{R}econfigurable intelligent surfaces (RIS) framework for RF \textbf{E}nergy harvesting. The UAV-mounted APs achieve significant performance in data harvesting for IoT devices due to their reachability and mobility~\cite{krijestorac2019uav}. The CFmMIMO UAVs can provide higher average signal strength even in the edge areas of the coverage region. Moreover, the RISs can reflect the incident signals into the desired direction with the controllable meta elements, which can effectively solve the non-line of sight problem and provide constructive interferences to the IoT devices. 

In summary, our contributions are three-fold:
\begin{itemize}
    \item We design and implement \name{}, a novel framework for improved RF energy harvesting with the combined benefits of CFmMIMO, UAV, and RIS.
    \item We verify our framework's performance by comparing it with the max-min fairness (MMF) algorithm proposed by Demir et al.~\cite{demir2020joint} with respect to spectral efficiency and harvested energy.
    \item We propose three deployment strategies for the RISs and make empirical comparisons to find out the optimal strategy.
\end{itemize}

\noindent\textit{Organization:} The rest of the paper is organized as follows: We discuss sufficient preliminary information in Section~\ref{sec:background}. The related works are discussed in Section~\ref{sec:related_works}. We introduce our proposed \name{} framework in Section~\ref{sec:proposed_framework}. In Section~\ref{sec:technical_details}, we discuss the technical details of the frameworks and the complete analysis of our algorithms. In Section~\ref{sec:evaluation}, we explain the evaluation setup along with the empirical analysis and findings. At last, we conclude the paper in Section~\ref{sec:conclusion}.

\section{Background}
\label{sec:background}
\begin{figure}[t]
    \centering
    \includegraphics[width=1\columnwidth]{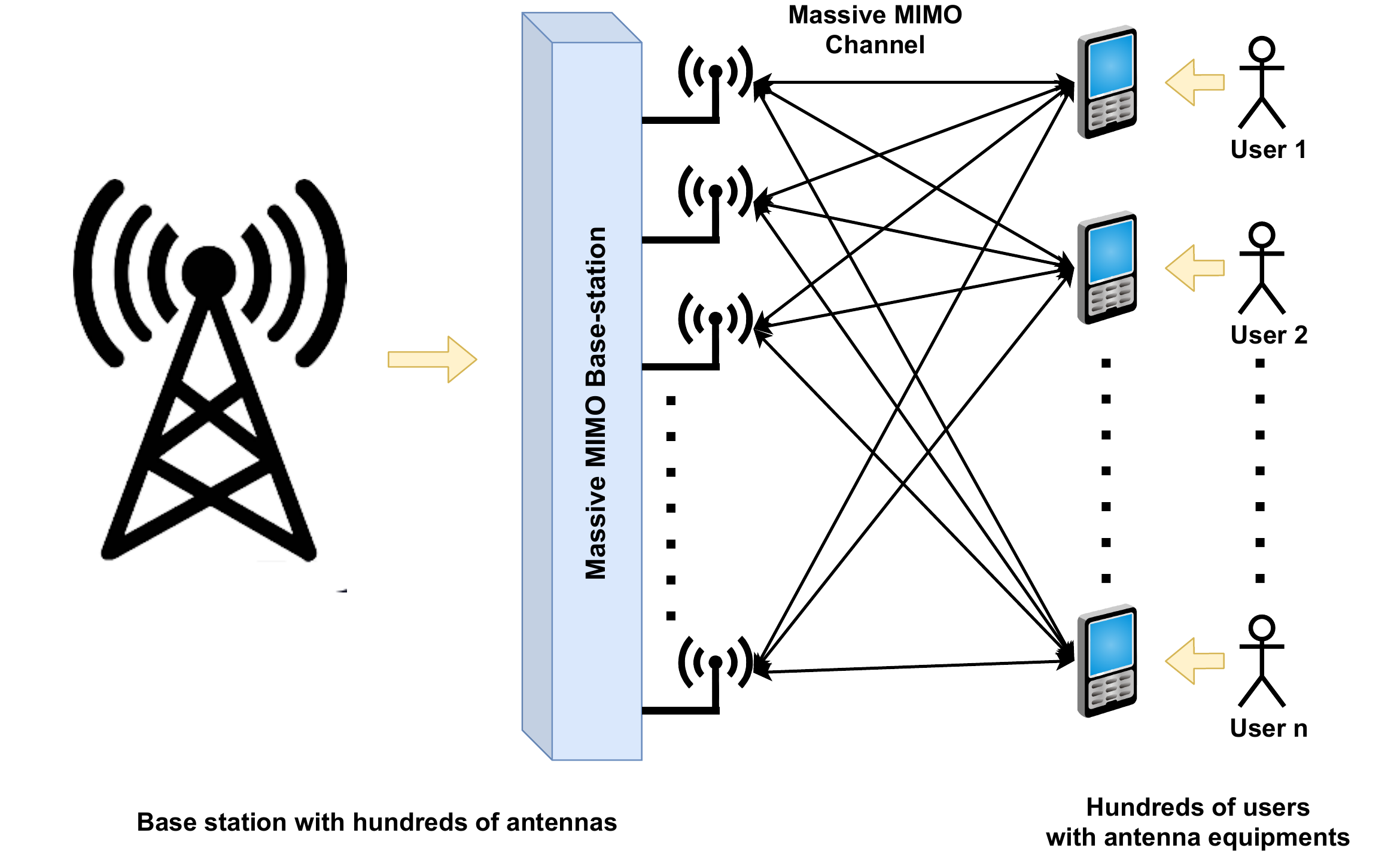}
    \caption{mMIMO architecture.}
    \label{fig:massmimo}
\end{figure}
In this section,  we discuss the necessary preliminary information that helps to explain the proposed \name{} mechanism.

\done{The standard is to follow "Capitalize Most Words" in the (section/subsection/any) titles. Please follow it.}

\subsection{Cell-Free Massive Multiple-input Multiple-output} \done{Why M in Massive is capital?}
The mMIMO has been receiving paramount interest throughout the last decade or so due to the high spectral efficiency (SE) it offers by the spatial multiplexing of large number of devices on the same time-frequency resource~\cite{wong2017key}. The basic architecture of the mMIMO is presented in Fig.~\ref{fig:massmimo}. Although mMIMO has been providing higher data transfer rates than preceding technologies, it suffers from inter-cell interference, especially for the cell-edge-devices~\cite{zhang2020prospective}. Moreover, the complexity of the symbol detector increases exponentially in the mMIMO uplink receiver, owing to the large number of antennas and RF chain~\cite{albreem2019massive}. As an alternative, more advantageous incarnation, CFmMIMO network infrastructure has been proposed by Ashikhmin et al.~\cite{ngo2017cell}, where a large number of antennas jointly can serve a lower number of IoT devices by relying on time-division duplex (TDD) operations. These antennas are geographically distributed throughout the coverage area and leverage the fronthaul network with CPU operating at the same time-frequency resource. The CPU sends power coefficients and downlink data to the APs, while the APs send back the uplink data received from the IoT devices, to the CPU via fronthaul link. The advantages to CFmMIMO include but not limited to high energy efficiency in terms of mbits/joule because of high array gain, flexible and cost-effective deployment, channel hardening and the favorable propagation conditions, appealingly uniform quality of service (QoS)~\cite{zhang2019cell}.

\begin{figure}[t]
    \centering
    \includegraphics[width=0.9\columnwidth]{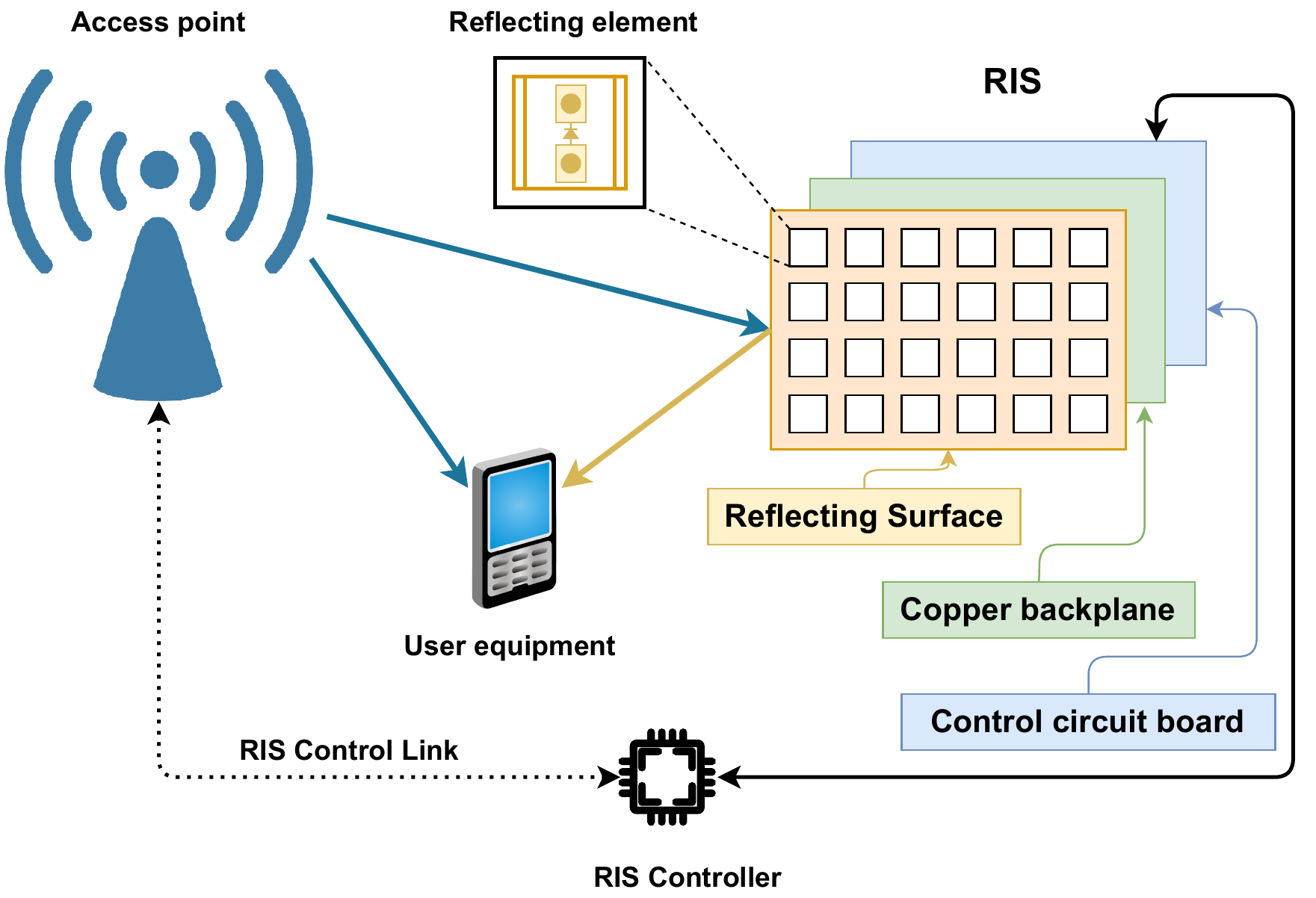}
    \caption{Reconfigurable intelligent surface architecture.}
    \label{fig:ris_arch}
\end{figure}

\subsection{Reconfigurable Intelligent Surfaces}

The RIS is a brand-new concept in the domain of wireless technologies that is drawing a lot of attention from the wireless research community. It is a 
relaying metasurface
\done{We always use metasurface than EM surface}, controlled with integrated electronics, and it brings unique wireless communication capabilities~\cite{basar2019wireless}. The basic architecture of RIS is presented in Fig.~\ref{fig:ris_arch}. It has a few layers of plane surface that can be fabricated with nano-printing and lithography techniques~\cite{huang2019reconfigurable}. The elements can independently provide some changes to the incident signal without consuming any bit of the transmit power. The change can be towards the phase, frequency, amplitude or even polarization~\cite{basar2019wireless}. The distinctive characteristic of RISs comes from the capability of controlling the environment via telecommunication operators that shape the EM response of the objects distributed throughout the network~\cite{liaskos2018new}. In simpler terms, when direct communication suffers bad qualities, the RIS (generally installed on walls, ceilings, and facades~\cite{zhao2019survey}) assists the transmission between sender and receiver by configuring the wireless environment.

\subsection{Unmanned Aerial Vehicle as Access Point}

\begin{figure}[t]
    \centering
    \includegraphics[width=0.92\columnwidth]{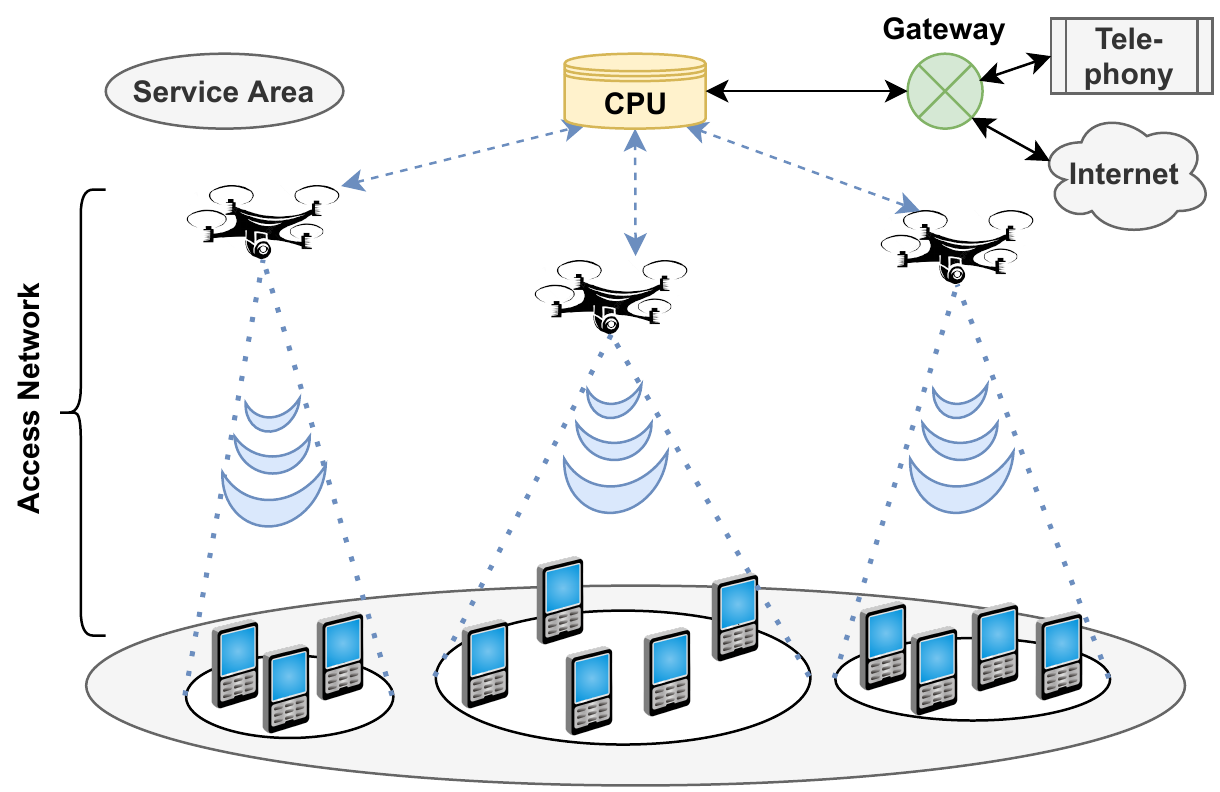}
    \caption{UAV-mounted Access Point.}
    \label{fig:uav_ap}
\end{figure}

The UAV technology is gaining a lot of interest recently due to its utilization in numerous
commercial and military applications~\cite{bushnaq2020optimal}. Because of the strong line-of-sight and flexible deployment of UAVs, they are becoming capable of assisting terrestrial wireless networks. Consequently, with UAVs being integrated with terrestrial cellular infrastructure, the next-generation network's QoS demand can be achieved~\cite{zeng2016wireless}. UAV-assisted cellular communication is specially fitted for providing extra coverage to `hot-spot' geographical regions like heavy traffic. Fig.~\ref{fig:uav_ap} illustrates the application of UAV-mounted APs, that communicate with the CPU through backhaul to provide cellular services in such coverage areas. However, there are few weaknesses of this technique, including the battery capacity and the service quality. The limited capacity of the state-of-the-art batteries will restrict the operational time of the UAV-mounted APs, while the QoS will be restricted by the capacity of the backhaul link between a fixed terrestrial base station (TBS) and the UAVs~\cite{bushnaq2020optimal}. A viable solution to these problems can the tethered UAVs, located on a mobile station or a rooftop, provided with power and data through cable from the TBS~\cite{kishk2020aerial}. 
The proposed model can either utilize tethered UAVs for long-lasting missions and untethered UAVs for time-limited missions.
\done{I recommend including the following sentence in the last paragraph: our model can either utilize tethered UAVs for long-lasting missions and untethered UAVs for time-limited missions.}

\section{Related Works}
\label{sec:related_works}

Downlink RF energy harvesting with mMIMO has become an active research topic in the scientific community in recent years. To mention some, Chen et al. investigated the maximization of energy efficiency for energy harvesting with mMIMO maintaining satisfiable QoS via energy beamforming~\cite{chen2013energy}. Amarasuriya et al. studied the performance of wireless energy transfer for multi-cell multi-way mMIMO relaying by deriving the harvested energy versus the achievable sum rate trade-offs~\cite{amarasuriya2016wireless}. Zhao et al. investigate the energy harvesting capability of mMIMO by maximizing the minimum harvested energy among the energy-requiring devices subject to a minimum achievable rate for information-requiring-devices~\cite{zhao2015downlink}. There are a lot of works on investigating energy harvesting using CFmMIMO as well. For example, Ngo et al. explore the performance of CFmMIMO with conjugate beamforming in downlink by considering the joint effect of power control, channel estimation, and nonorthogonality of pilot sequences~\cite{ngo2017cell}. Nayebi et al. investigate the downlink performance of CFmMIMO with zero-forcing (ZF) precoding and conjugate beamforming, and propose an algorithm with low complexity power allocation~\cite{nayebi2017precoding}. Unlike the above-mentioned works, Shrestha et al. investigate the performance of simultaneous wireless information and power transfer (SWIPT) for training-based CFmMIMO~\cite{shrestha2018swipt}. However, they considered that information and energy users are located separately. Likewise, Alageli et al.~\cite{alageli2019optimal} studied SWIPT with CFmMIMO where only the energy users harvest energy, and the information users don't. Following the methodology of~\cite{demir2020joint}, we consider energy harvesting of all the users regardless of their primary requirement by considering power control for maximizing the minimum uplink spectral efficiency for wireless power transfer with CFmMIMO. Moreover, we introduce the idea of leveraging RIS assisting the CFmMIMO mounted on the tethered UAV APs for improved energy harvesting performance. To the best of our knowledge, this is the first work that combines these three technologies for RF energy harvesting.

\section{Proposed Framework}
\label{sec:proposed_framework}

\begin{figure}[hbt!]
    \centering
    \includegraphics[width=1\columnwidth]{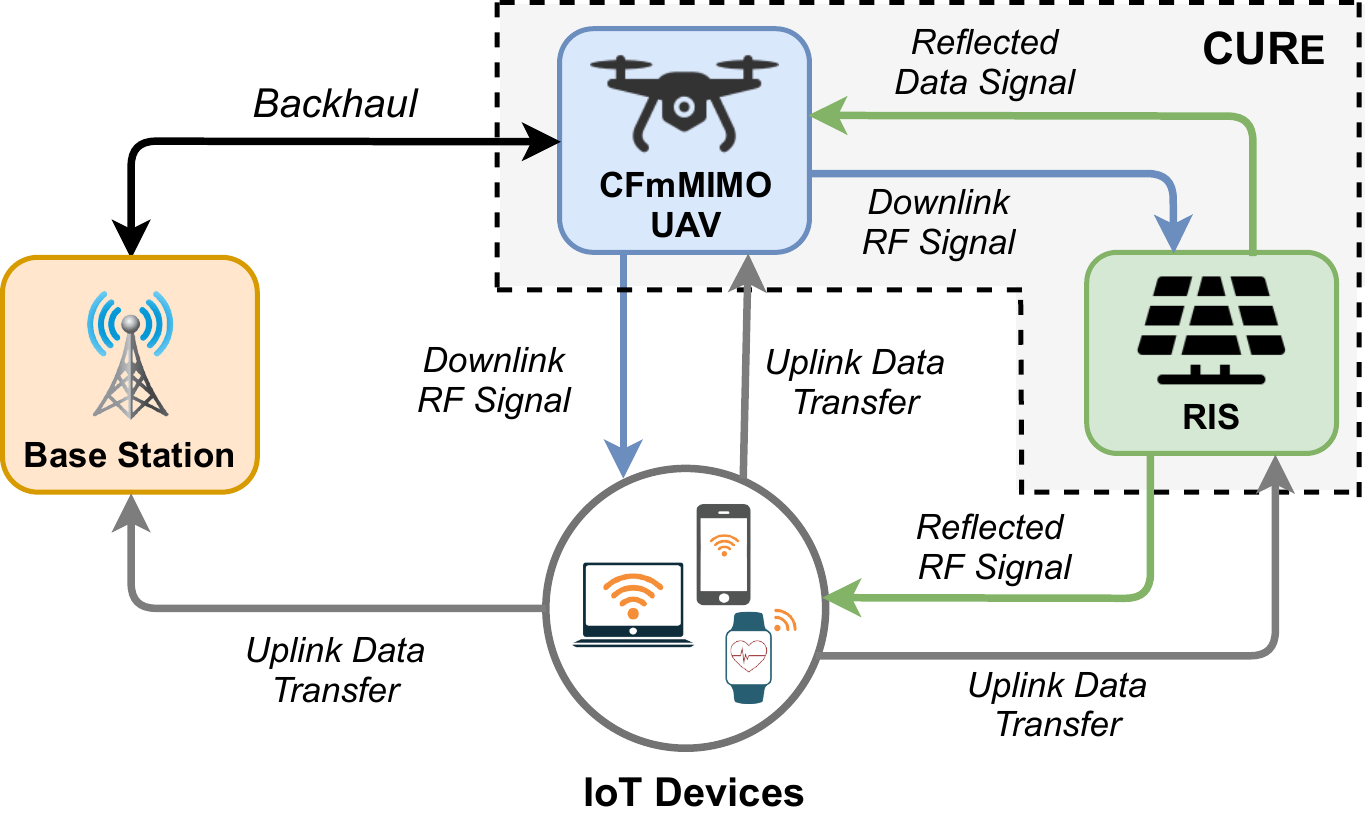}
    \caption{Proposed \name{} framework.}
    \label{fig:fw}
\end{figure}

We discuss our proposed \name{} framework in this section. As presented in Fig.~\ref{fig:fw}, our framework provides a downlink radio frequency signal to the battery-limited IoT devices for energy harvesting. The UAVs, mounted with cell-free mMIMO, act as the access points. We consider that the UAVs are tethered with a direct energy source. So, they hover at a point for acting as the access points. They also have an immediate link with the core network, i.e., base station. They provide direct RF signals to the IoT devices \done{I recommend using IoT devices than IoT networks. If you are going to go with this recommendation, then change it in the figure also}. 

\done{for the paragraph below: RIS is not used mainly to enhance the coverage for the cell edge. It is enhancing the EH in the downlink and the data transmission in the uplink.}
On the other hand, for enhancing the energy harvesting in the downlink and for providing signals to the non-line-of-sight locations in the coverage areas, reconfigurable intelligent surfaces are utilized. They reflect the RF signal intelligently towards the IoT devices. The combined RF signal from the access points and RISs are harvested at the user end. Then, this power is leveraged for the uplink information transfer from the user devices to the access points. 
Depending on the location, the uplink signal can be sent directly to either the base station or the UAV/AP and/or via the RIS panel. 
\done{Depending on the location, the uplink signal can be sent directly to the UAV/AP and/or via the RIS panel.}

\section{Technical Details}
\label{sec:technical_details}
In this section, we discuss the technical details of our proposed \name{} framework. First, we discuss the underlying system model. Later, we explain our downlink energy harvesting approach. Finally, we provide a detailed description of Algorithm~\ref{algo}.

\subsection{System Model}
We consider cell-free mMIMOs mounted on the tethered UAVs as the access points. For serving $J$ number of IoT devices with harvesting capability across the coverage area, we assume $U$ number of APs are distributed. Each UAV-mount CFmMIMO is assumed to be equipped with $N$ number of antennas and they have an error-free fronthaul connection to the central processing unit (CPU). We leverage the implementation scheme proposed by Demir et al.~\cite{demir2020joint}. Accordingly, we assume a time division duplex (TDD) operation, which will force channel reciprocity. We denote the total number of samples per coherence interval by $\delta_c$. Each of the coherent intervals is split into three phases: (i) uplink training, (ii) downlink wireless power transfer and (iii) uplink wireless information transfer. In the first phase, all the IoT devices send pilot sequences of length $\delta_p$ to the UAV APs, for estimating the channel in order to design the precoding vectors for efficient energy transfer and data reception. For the downlink and uplink transfer, $\delta_d$ and $\delta_u$ samples are used respectively. So, for each coherent interval, the total samples:
\begin{equation}
  \mathcal{\delta}_{c} = \mathcal{\delta}_{p} + \mathcal{\delta}_{d} + \mathcal{\delta}_{u}
\end{equation}

We denote the channel between the $j^{th}$ user and the $u^{th}$ AP by $\textbf{h}_{ju}$, where the channels remain constant in a single time-frequency coherence interval. In the context of CFmMIMO UAVs with multiple antennas, we follow the spatially uncorrelated Rician fading channels proposed by Demir et al.~\cite{demir2020joint} with unknown phase shifts. So the realization of each channel can be expressed as:
\begin{equation}
    \textbf{h}_{ju} = e^{k{\phi}_{ju}} \bar{\textbf{h}}_{ju} + \Tilde{\textbf{h}}_{ju}
\end{equation}

Here, $e^{k{\phi}_{ju}} \bar{\textbf{h}}_{ju}$ and $\Tilde{\textbf{h}}_{ju}$ represents the line-of-sight (LOS) and non-line-of-sight (NLOS) components respectively. For NLOS component, the small-scale fading is modeled as $\mathcal{N}_{\mathcal{C}}(\textbf{0}_\textit{N}, \gamma_{ju} \textbf{I}_\textit{N})$, with $\gamma_{ju}$ representing the large-scale fading co-efficient. In accordance with prior literature, we consider that the UAV APs have perfect knowledge about the LOS component and large-scale fading co-efficient corresponding to the channel between the IoT devices and themselves, describing the long-term channel effects. Unlike the former works that consider the negligence of phase shift $\phi_{ju}$ by the Rician fading, we consider the realistic scenario where, due to user mobility, $\phi_{ju}$ in the LOS component is unknown. A little amount of random $\phi_{ju}$ is induced on both the LOS and NLOS component, constructed by the individual paths, when receiver and transmitter move over distances at the order of the wavelength.

\subsection{Downlink Energy Harvesting}

\begin{figure}[t]
    \centering
    \includegraphics[width=.9\columnwidth]{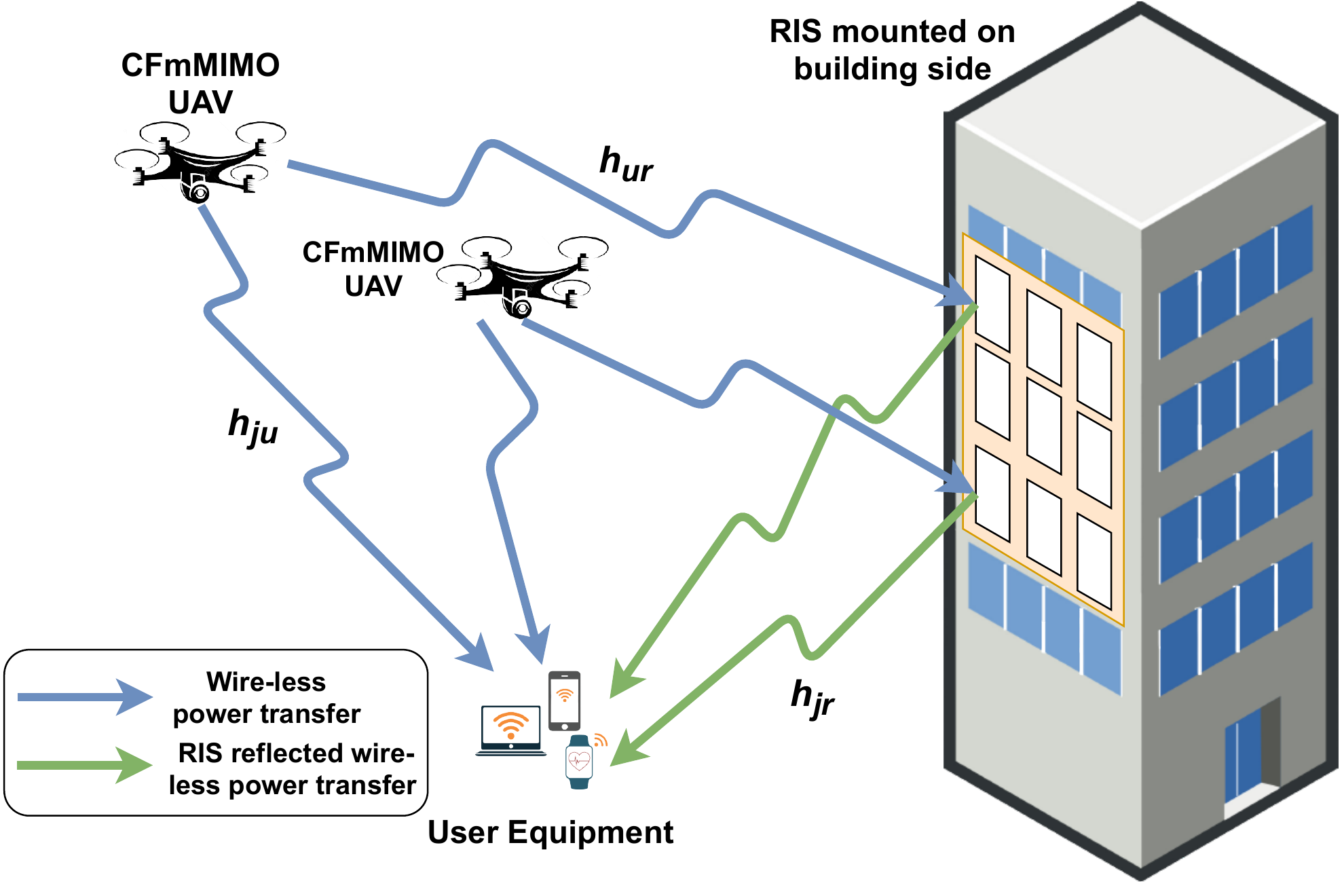}
    \caption{Proposed RIS assisted wire-less power transfer.}
    \label{fig:ris_assist}
\end{figure}

In this phase, all APs start to transmit energy to the IoT devices by utilizing the channel state information (CSI) for the downlink precoding. First, the coherent energy transmission will be analyzed where same energy symbol is transmitted by all the APs for each IoT devices in a synchronous
manner, to increase the harvested energy. The signal transmitted by the $u^{th}$ AP can be expressed as:
\begin{equation}
    \textbf{x}_{u}^{E} = \sum_{j=1}^{J} \sqrt{p_{ju}} \textbf{w}_{ju}^{*}s_j
\end{equation}

Here, $\textbf{w}_{ju}^{*}$ denote the downlink precoding vector for this phase. $s_j$ and $p_{ju}$ represent the zero-mean unit-variance energy signal for the $j_{th}$ IoT device and the power control
coefficient of the $u_{th}$ AP corresponding to the $j_{th}$ device. In the long term, the transmission power for each CFmMIMO AP should satisfy the maximum power limit:
\begin{equation}
    P_{u}^{E} \triangleq \mathrm{E}  \bigg\{ {\left\|  \textbf{x}_u^E \right\|}^2 \bigg\} \leq \rho_d
\end{equation}

Here, $P_{u}^{E}$ is the average transmitted power for the $u_{th}$ AP and $\rho_d$ is the maximum power limit. Again $P_{u}^{E}$ can be calculated by:
\begin{equation}
    P_{u}^{E} = \mathrm{E}  \Bigg\{ {\left\|  \sum_{j=1}^{J} \sqrt{p_{ju}} \textbf{w}_{ju}^{*}s_j \right\|}^2 \Bigg\} = \sum_{j=1}^{J} p_{ju} \mathrm{E}  \bigg\{ {\left\|  \textbf{w}_{ju} \right\|}^2 \bigg\}
\end{equation}

The received signal for the $j_{th}$ IoT device is:
\begin{equation}
    r_{j}^{E} = \sum_{u=1}^{U} \textbf{h}_{ju}^T \textbf{x}_{u}^{E} + n_{j}^{E} = \sum_{u=1}^{U} \sum_{m=1}^{J} \sqrt{p_{mu}} \textbf{w}_{mu}^{H} \textbf{h}_{ju} s_m + n_{j}^{E}
\end{equation}

Here, $n_{j}^{E}$ indicated the additive noise at the $j_{th}$ IoT device. The average input power at the energy harvesting rectifier circuit of the $j_{th}$ device can be expressed as:
\begin{equation}
    I_{j} = \mathrm{E}  \bigg\{ {\left| \sum_{u=1}^{U} \sum_{m=1}^{J} \sqrt{p_{mu}} \textbf{w}_{mu}^{H} \textbf{h}_{ju} s_m \right|}^2 \bigg\}
\end{equation}

Similar to~\cite{chen2017wireless}, we will utilize the following non-linear energy harvesting model, as this model correlates with real measured data. For the $j_{th}$ IoT device in $\delta_d$ channel, the total harvested energy can be expressed as:
\begin{equation}
    E_{j} = \frac{\delta_d A_j I_j}{B_j I_j + C_j}
\end{equation}

here, $A_j$ $>$ 0, $B_j$ $\geq$ 0, and $C_j$ are constants determined by curve fitting of the rectifier circuit of the $j_{th}$ device~\cite{chen2017wireless}. 

\done{The equations and notations above this remark is totally disconnected from those below this remark. For example, above you used the channel gain symbol as g. Below, you used the channel gain symbol as h. Please try to make sure that all the equations are consistent.}

For the RIS supported transmission, we leverage the expressions from~\cite{bjornson2019intelligent}. Fig.~\ref{fig:ris_assist} presents a detailed view of the RIS assisted energy harvesting. The RIS has N discrete elements and the deterministic channel from source to RIS is presented by $\textbf{h}_{ur}$ ($n_{th}$ component is presented by $[\textbf{h}_{ur}]_n$). The channel in between the destination and the RIS is represented by $\textbf{h}_{jr}$ and the deterministic flat-fading
channel is denoted by $\textbf{h}_{ju}$, as mentioned in the earlier equations. The RIS properties are represented by:
\begin{equation}
    \Theta = \alpha diag(e^{j\theta_1},..., e^{j\theta_N})
\end{equation}

here, $\Theta$ represents the diagonal matrix, $\alpha$ $\in$ (0,1] and $\theta_1,...,\theta_N$ represent the fixed amplitude reflection coefficient and the phase-shift variables respectively. The received signal at the destination can be expressed as:
\begin{equation}
    S_r = (\textbf{h}_{ju}+\textbf{h}_{ur}^T \Theta \textbf{h}_{jr})\sqrt{xy}+n
\end{equation}

here,  $x, y,$ and $n$ represent the transmit power, unit-power information signal and receiver noise respectively. The channel capacity of the RIS-supported network can be expressed as:
\begin{align}
    R_{RIS}(N) = \max_{\theta_1,...,\theta_N} \log_2 \left( 1 + \frac{x {|\textbf{h}_{ju}+\textbf{h}_{ur}^T \Theta \textbf{h}_{jr}|}^2}{\sigma^2} \right)\\
    = \log_2 \left( 1 + \frac{x {(|\textbf{h}_{ju}|+ \alpha \sum_{n=1}^{N} |[\textbf{h}_{ur}]_n [\textbf{h}_{jr}]_n|)}^2}{\sigma^2} \right)
\end{align}

The maximum rate is achieved when the phase-shifts are set as $\theta_n$ = $arg(\textbf{h}_{ju})-arg([\textbf{h}_{ur}]_n [\textbf{h}_{jr}]_n)$. For brevity, the above equation can be re-written with: 
\begin{align}
    |h_{ju}| = \sqrt{\beta_{ju}}, |h_{ur}| = \sqrt{\beta_{ur}}, |h_{jr}| = \sqrt{\beta_{jr}}\\
    \frac{1}{N} \sum_{n=1}^{N} |[\textbf{h}_{ur}]_n [\textbf{h}_{jr}]_n| = \sqrt{\beta_{RIS}}
\end{align}

The re-written equation would be:
\begin{align}
    R_{RIS}(N) = \log_2 \left( 1 + \frac{x {(\sqrt{\beta_{ju}}+ N\alpha \sqrt{\beta_{RIS}})}^2}{\sigma^2} \right)
\end{align}

We utilize this equation for calculating the rates for our RIS assisted energy harvesting.
\begin{algorithm}[!t]
\DontPrintSemicolon
$Param_{sys}$ = [$Realizations, Pow_{transmit}, Pow_{noise},$\\ 
\qquad \qquad \quad \quad $Block_{coher}, Param_{rectifier}, Carrier$];\\
$Param_{sys}$ $\leftarrow$ Initialize;\\
$AllSetups$ $\leftarrow$ Number of setups;\\
$Matrix_{rate}$ $\leftarrow$ $\phi$;\\
\For{each $setup$ $\in$ $AllSetups$}
    {
        $rates$ $\leftarrow$ $\phi$;\\
        \While{TRUE}
        {
            $Gain_{Channel}, Realization_{Channel}$ $\leftarrow$ $\textbf{SetupFunc}$($Param_{AP}, Param_{RIS}$);\\
            $StatTerms, HarvestedEnergy$ $\leftarrow$
            $\textbf{ChannelEstimation}$($Gain_{Channel}$,\\ \qquad \qquad $Realization_{Channel}$, $Param_{sys}$);\\
            $rates$ $\leftarrow$ $\textbf{SpectralEfficiency}$($StatTerms$,\\ \qquad \qquad \qquad
            $HarvestedEnergy$, $\beta$);\\
            $Solution$ $\leftarrow$ $\textbf{Feasibility}$($rates$);\\
            \If{$Solution$ is $Feasible$}
            {
                $Matrix_{rate}$.append($rates$);\\
                break;\\
            }
            \Else
            {
                continue;\\
            }
        }
    }
\normalsize
\caption{Harvested Energy}
\label{algo}
\end{algorithm}

\begin{figure*}[t]
    \begin{center}
    \hspace{-10pt} 
     \subfigure[]
        {
        \label{fig:cdf_mmf1}
            \includegraphics[width=0.245\textwidth, keepaspectratio=true]{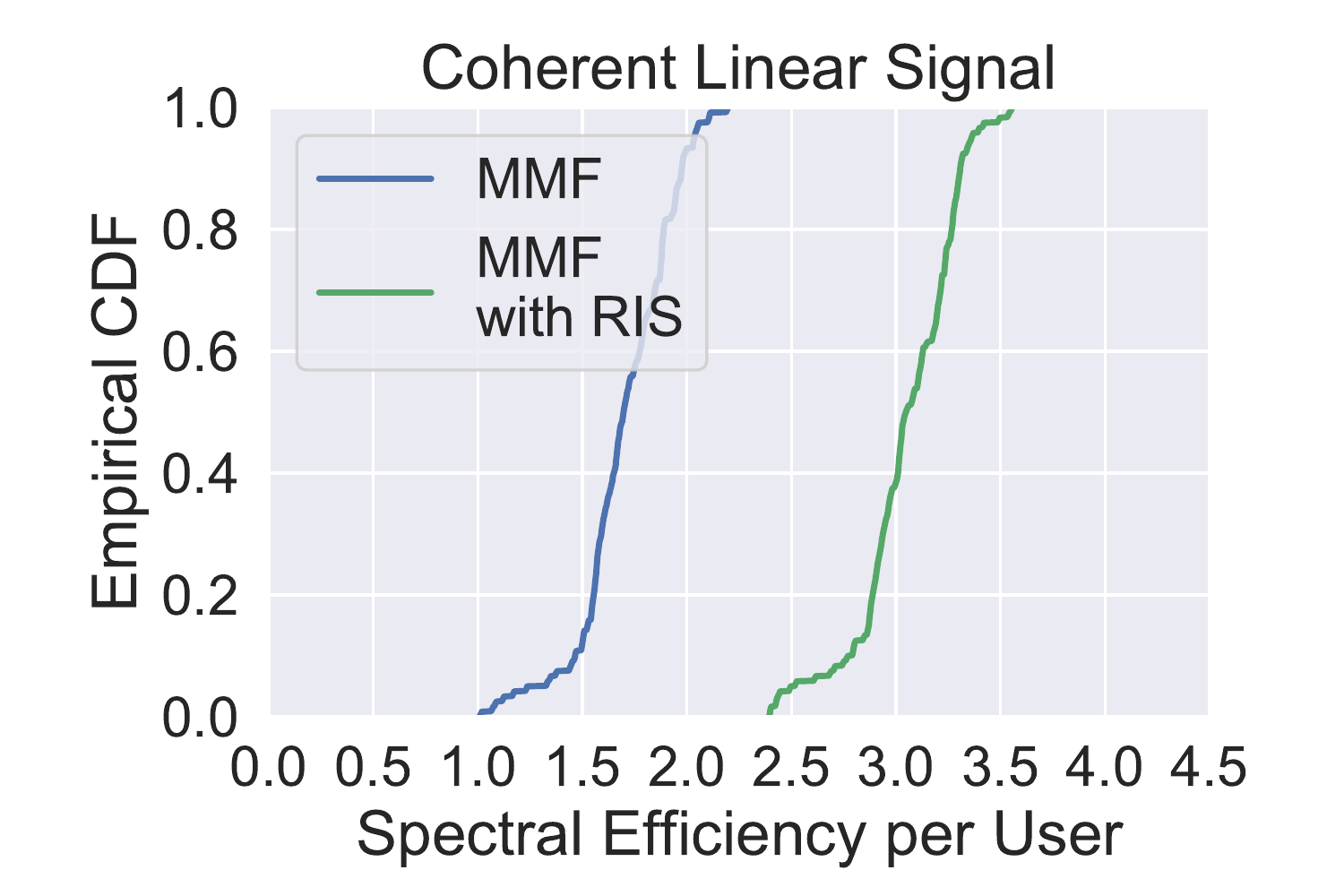}
        }
        \hspace{-10pt} 
    \subfigure[]
        {
        \label{fig:cdf_mmf2}
            \includegraphics[width=0.245\textwidth, keepaspectratio=true]{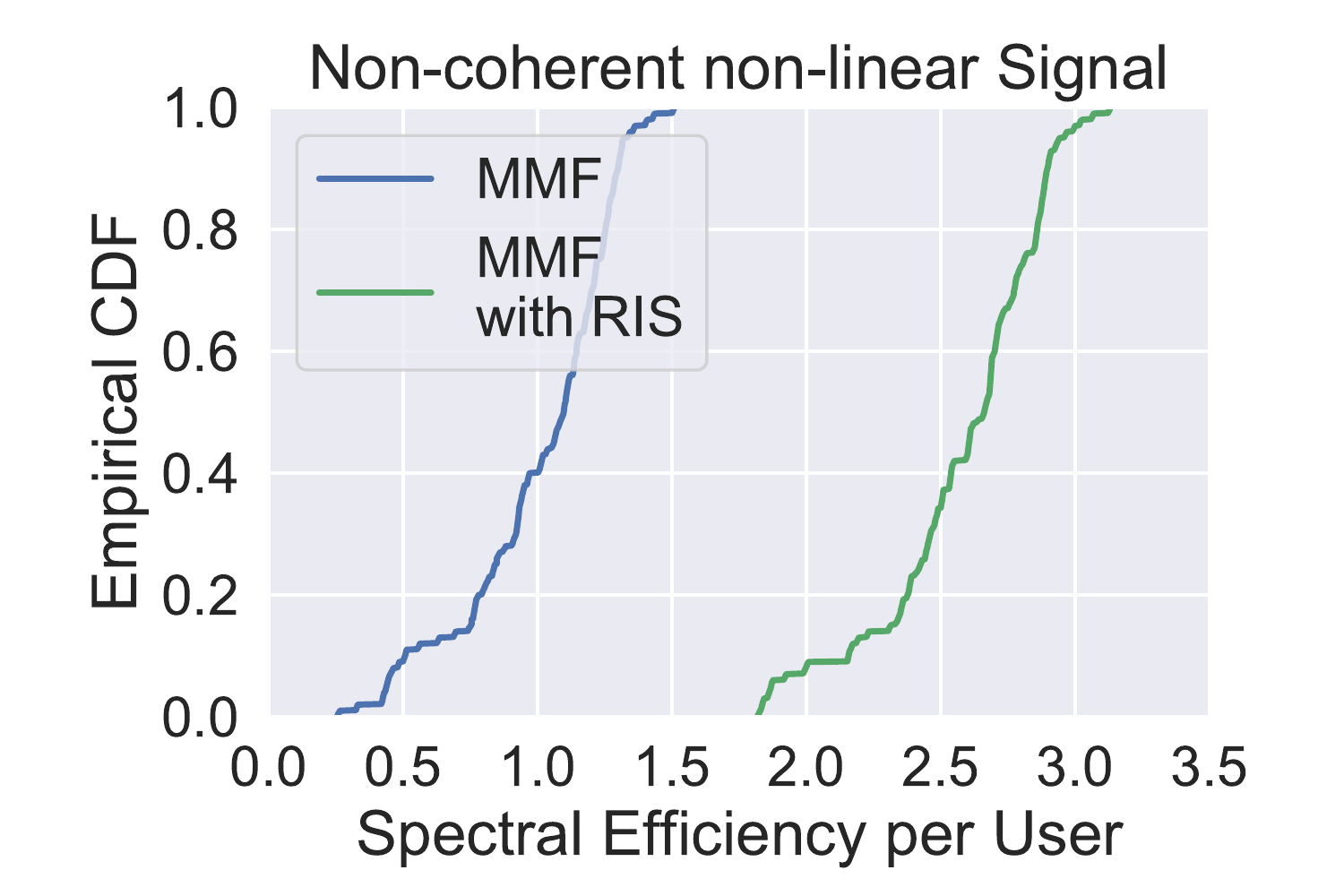}
        }
        \hspace{-10pt} 
    \subfigure[]
        {
        \label{fig:cdf_mmf3}
            \includegraphics[width=0.245\textwidth, keepaspectratio=true]{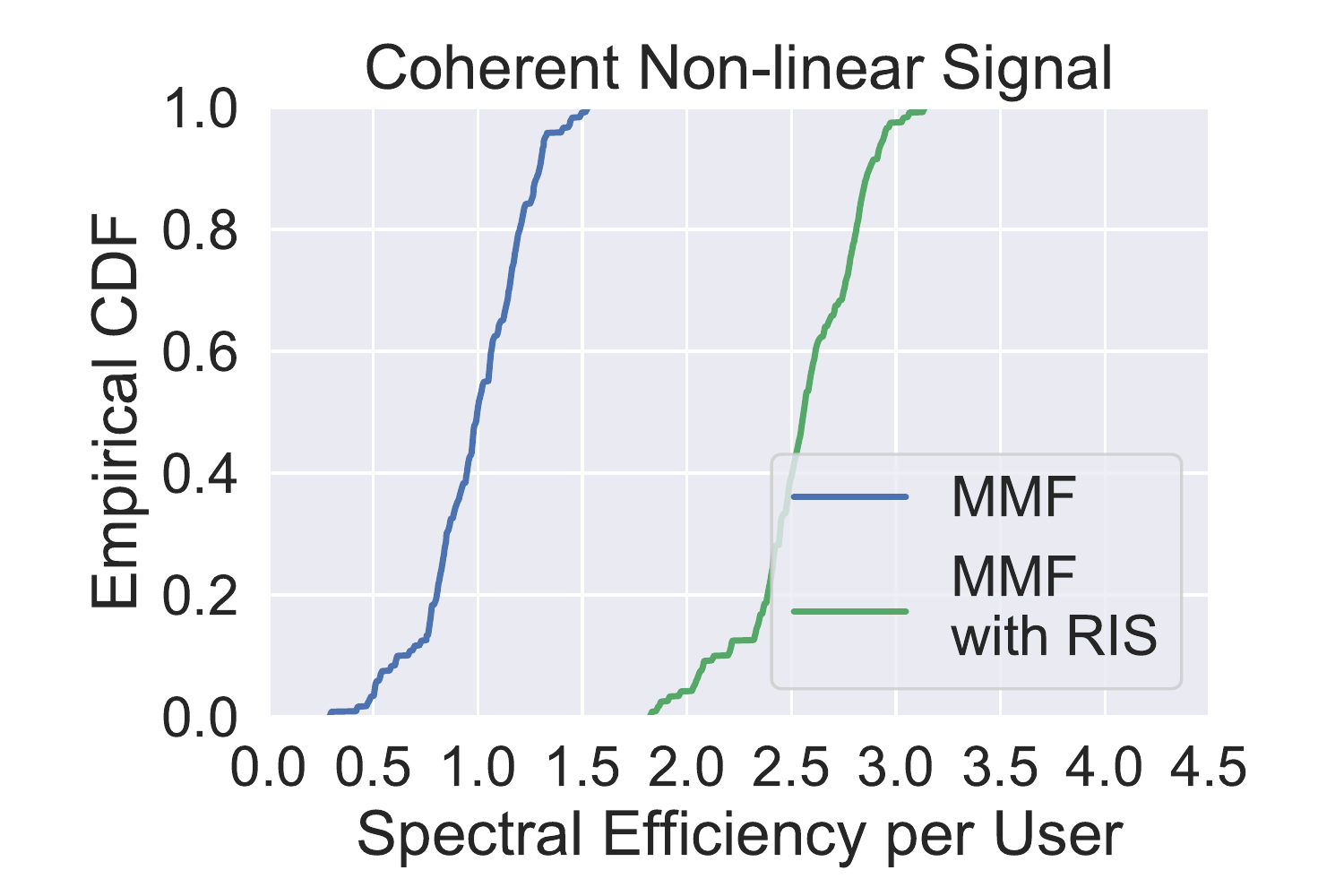}
        }
        \hspace{-10pt}
    \subfigure[]
        {
        \label{fig:cdf_mmf4}
            \includegraphics[width=0.245\textwidth, keepaspectratio=true]{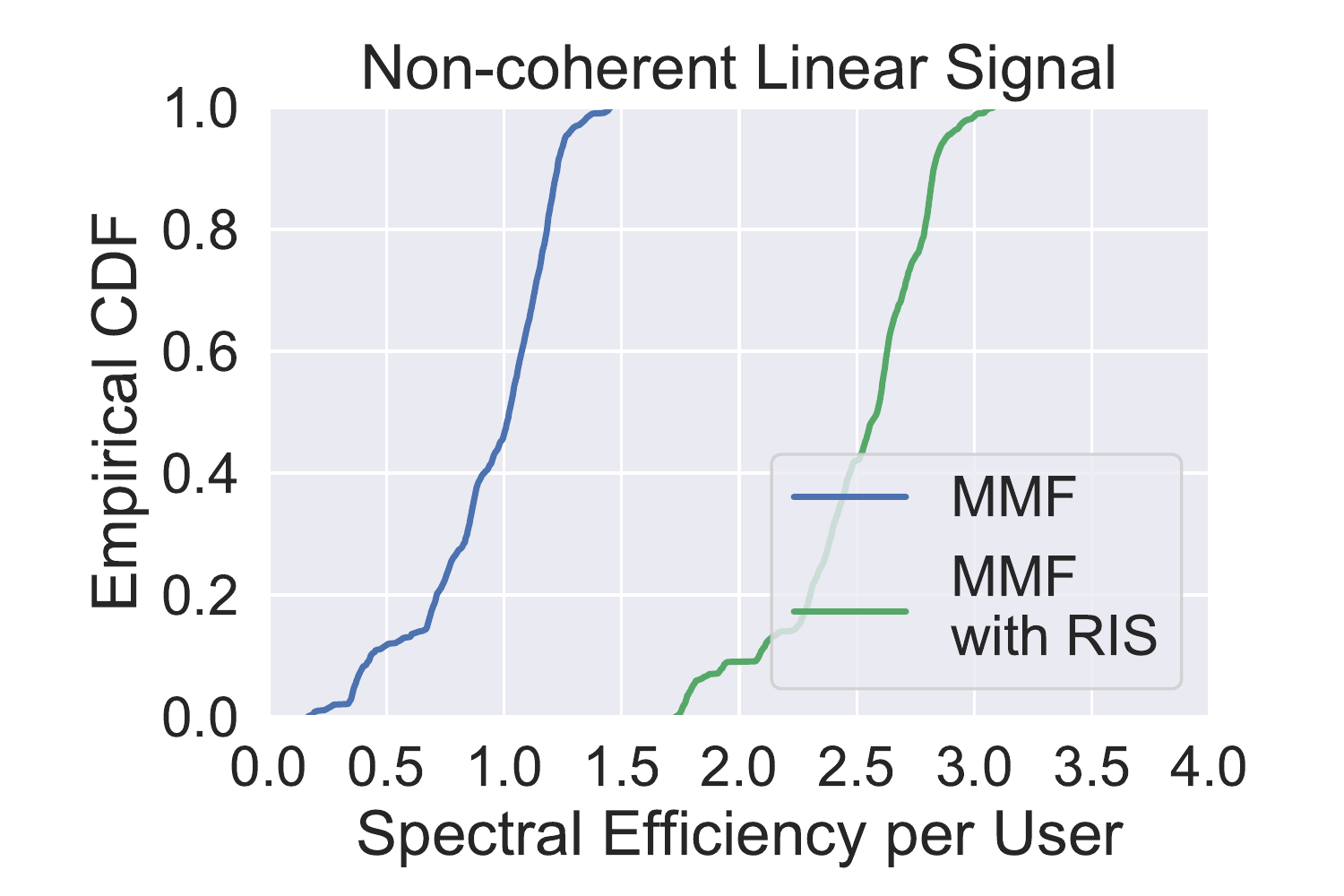}
        } 
    \end{center}
    \vspace{-15pt}
    \caption{Empirical cumulative distributive function of spectral efficiency per user for (a) coherent linear signal, (b) non-coherent non-linear signal, (c) coherent non-linear signal and (d) non-coherent linear signal, for 100 randomly deployed user equipment setups. \rahman{These figures are unnecessarily large than the information it is demonstrating. Try to have graphs, if possible, to fit 3 or 4 in a row.} \alvi{for this figure, I can have 2 more graphs: coherent non-linear and  non-coherent linear. }} 
    \label{fig:cdf_mmf}
    \vspace{-9pt}
\end{figure*}

\begin{figure*}[t]
    \begin{center}
    \hspace{-10pt} 
     \subfigure[]
        {
        \label{fig:nap1}
            \includegraphics[width=0.23\textwidth, keepaspectratio=true]{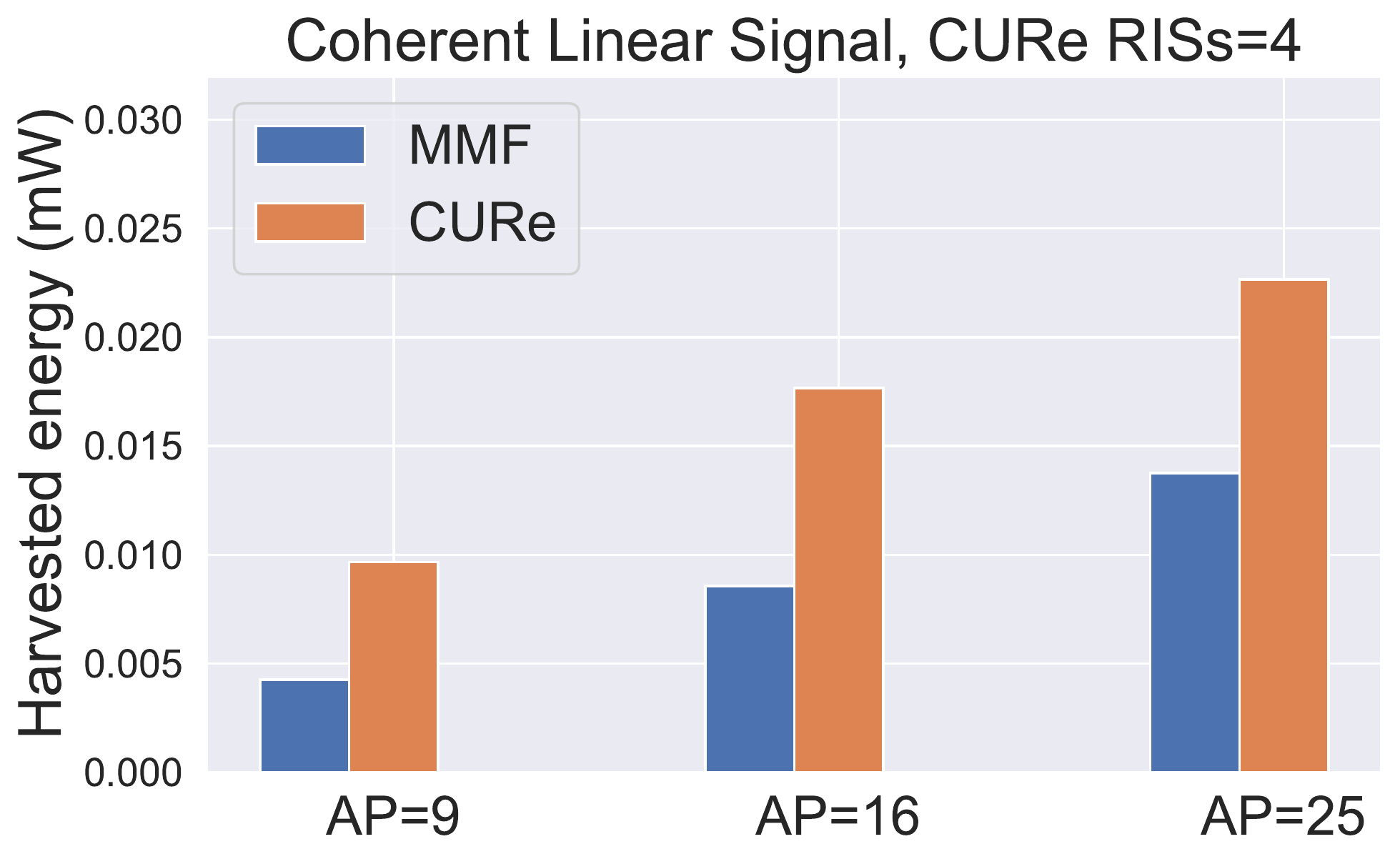}
        }     
        \hspace{-5pt} 
        \subfigure[]
        {
        \label{fig:nap2}
            \includegraphics[width=0.23\textwidth, keepaspectratio=true]{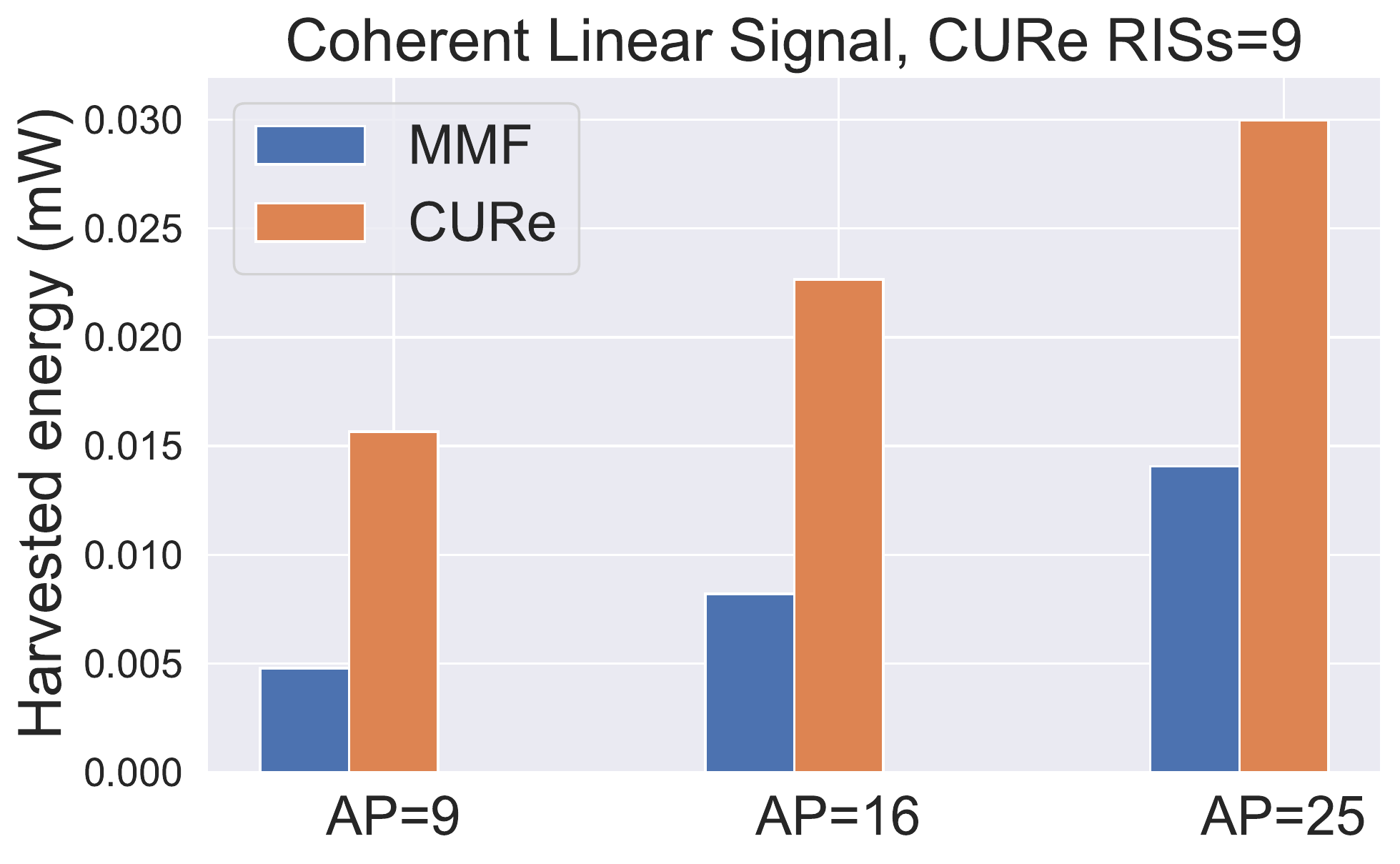}
        }
        \hspace{-5pt} 
        \subfigure[]
        {
        \label{fig:nap3}
            \includegraphics[width=0.23\textwidth, keepaspectratio=true]{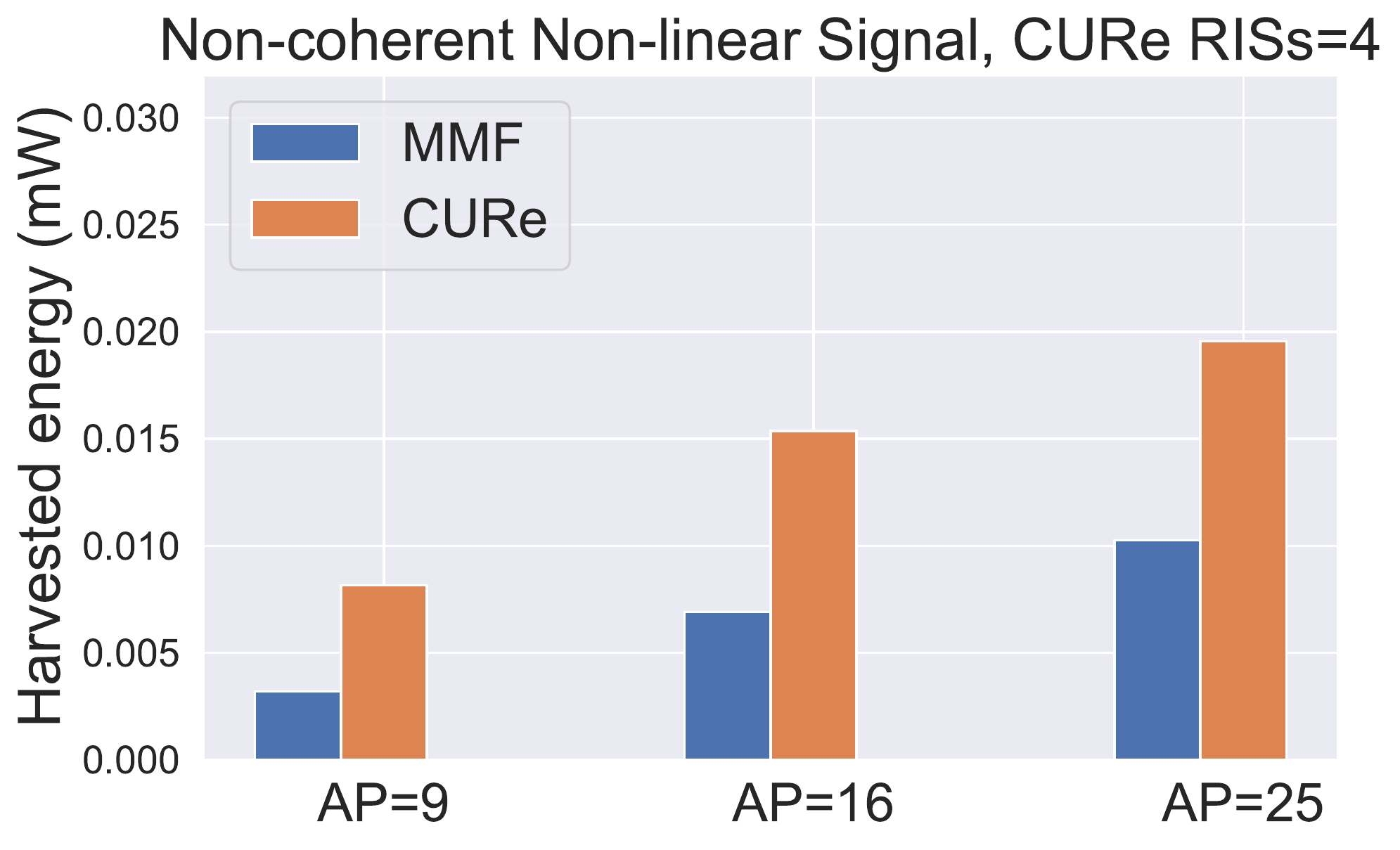}
        }
        \hspace{-5pt} 
        \subfigure[]
        {
        \label{fig:nap4}
            \includegraphics[width=0.23\textwidth, keepaspectratio=true]{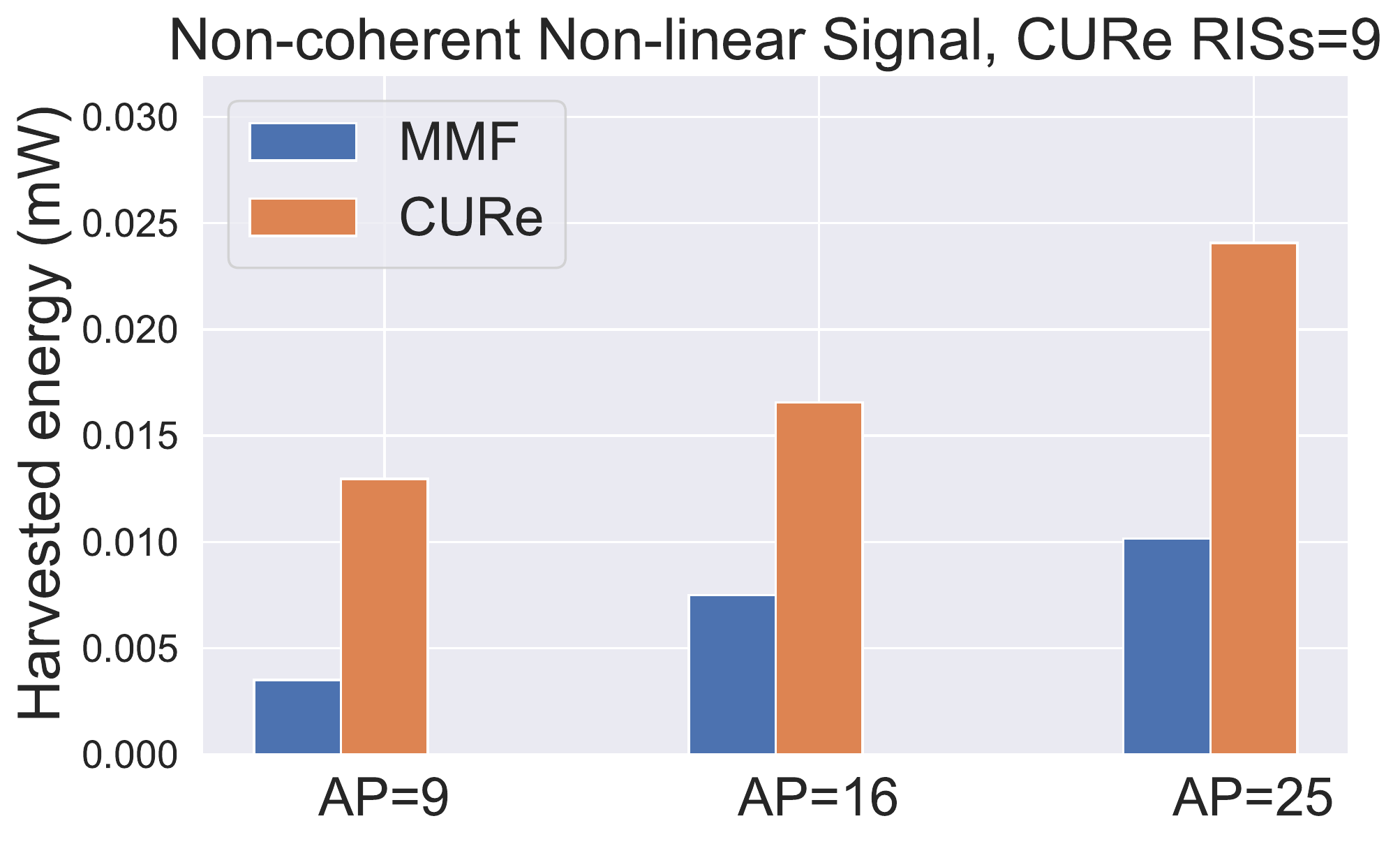}
        }
      
    \end{center}
    \vspace{-15pt}
    \caption{Comparison of MMF and propsed \name{} with respect to harvested energy for (a) Coherent Linear signal with 4 RISs for \name{}, (b) Coherent Linear signal with 9 RISs for \name{}, (c) Non-coherent Non-linear signal with 4 RISs for \name{}, (d) Non-coherent Non-linear signal with 9 RISs for \name{}.} 
    \label{fig:nap}
    \vspace{-9pt}
\end{figure*}

      

\begin{figure*}[t]
    \begin{center}
    \hspace{-10pt} 
     \subfigure[]
        {
        \label{fig:elements1}
            \includegraphics[width=0.23\textwidth, keepaspectratio=true]{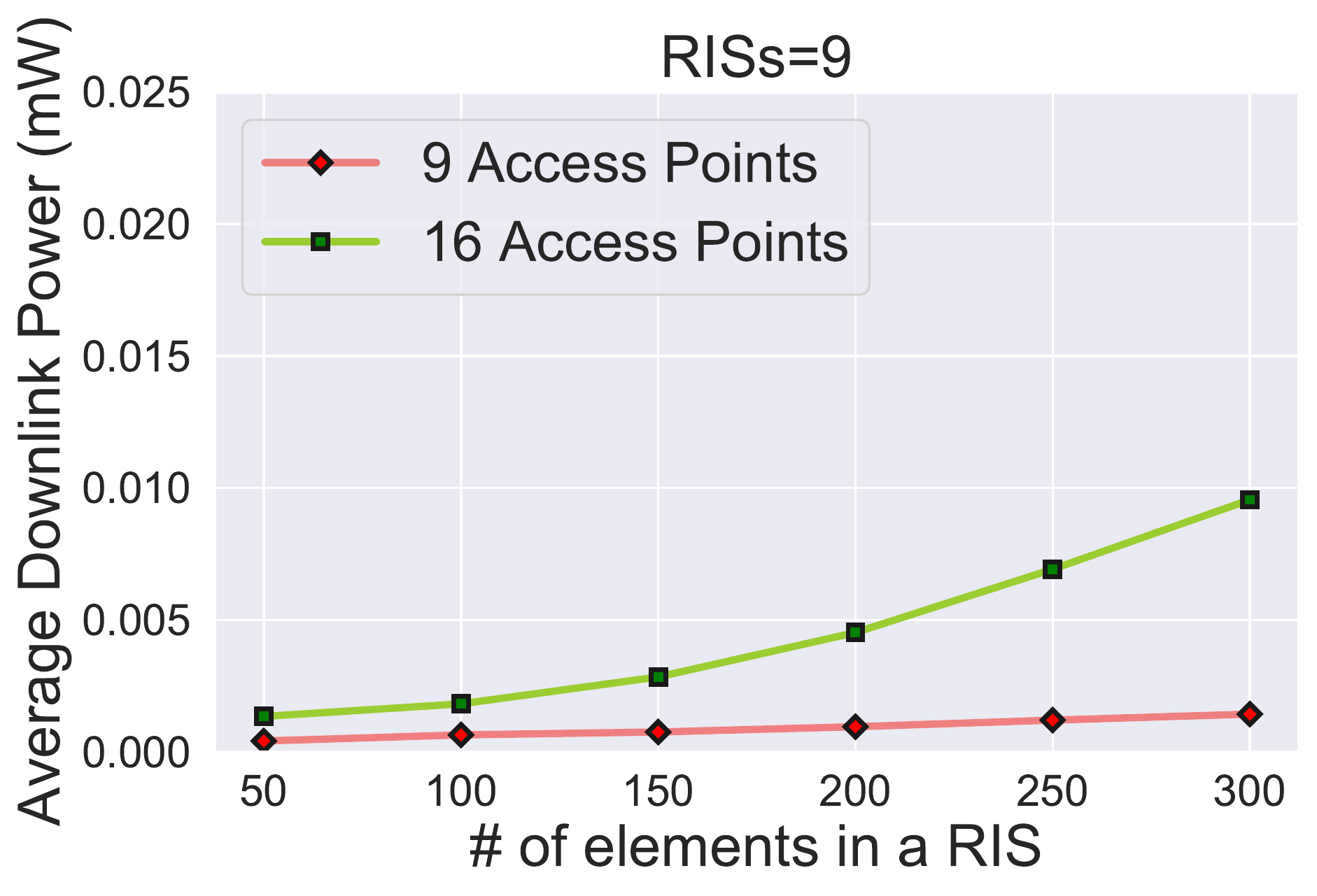}
        }     
        \hspace{-5pt} 
        \subfigure[]
        {
        \label{fig:elements2}
            \includegraphics[width=0.23\textwidth, keepaspectratio=true]{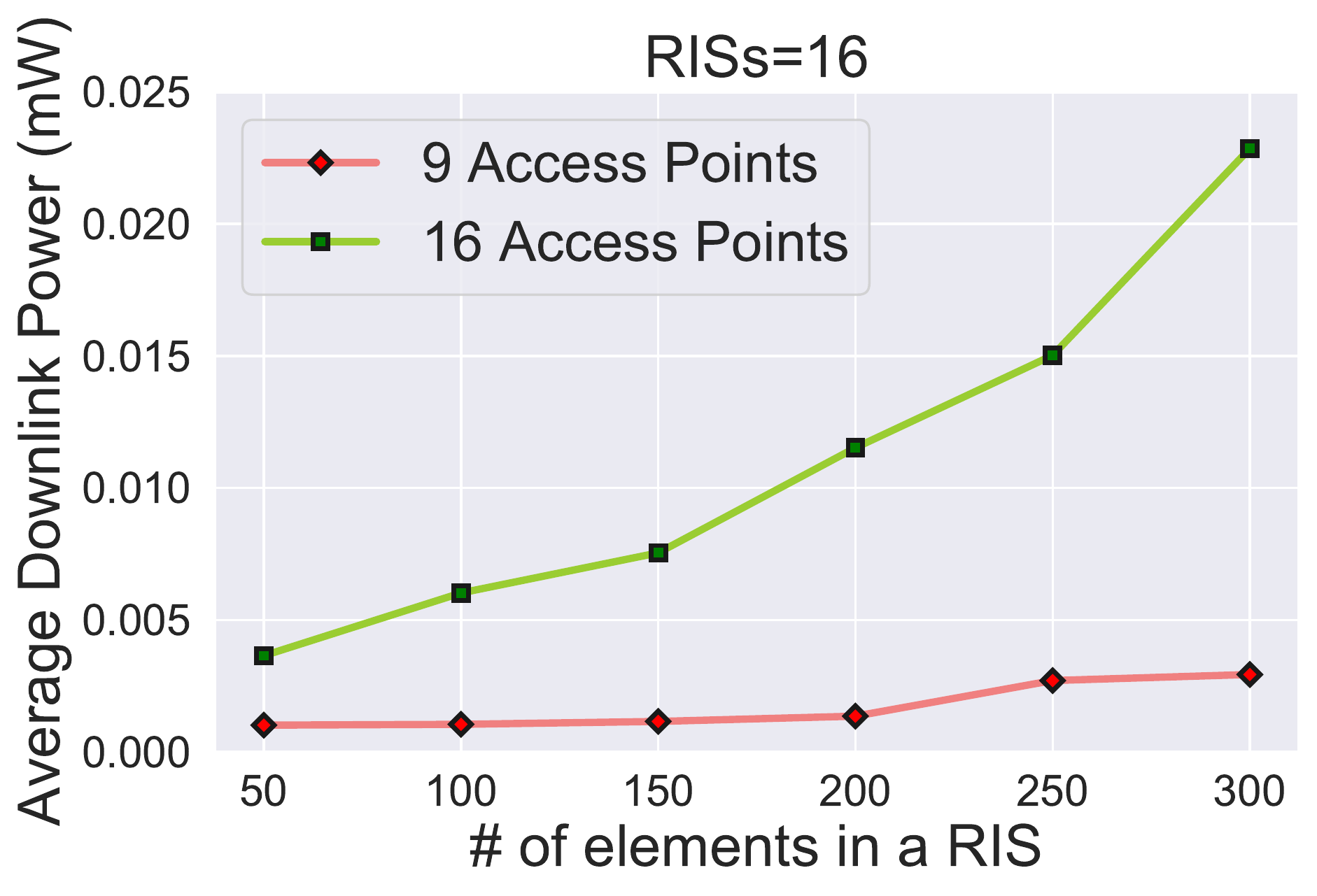}
        }
        \hspace{-5pt} 
        \subfigure[]
        {
        \label{fig:elements3}
            \includegraphics[width=0.23\textwidth, keepaspectratio=true]{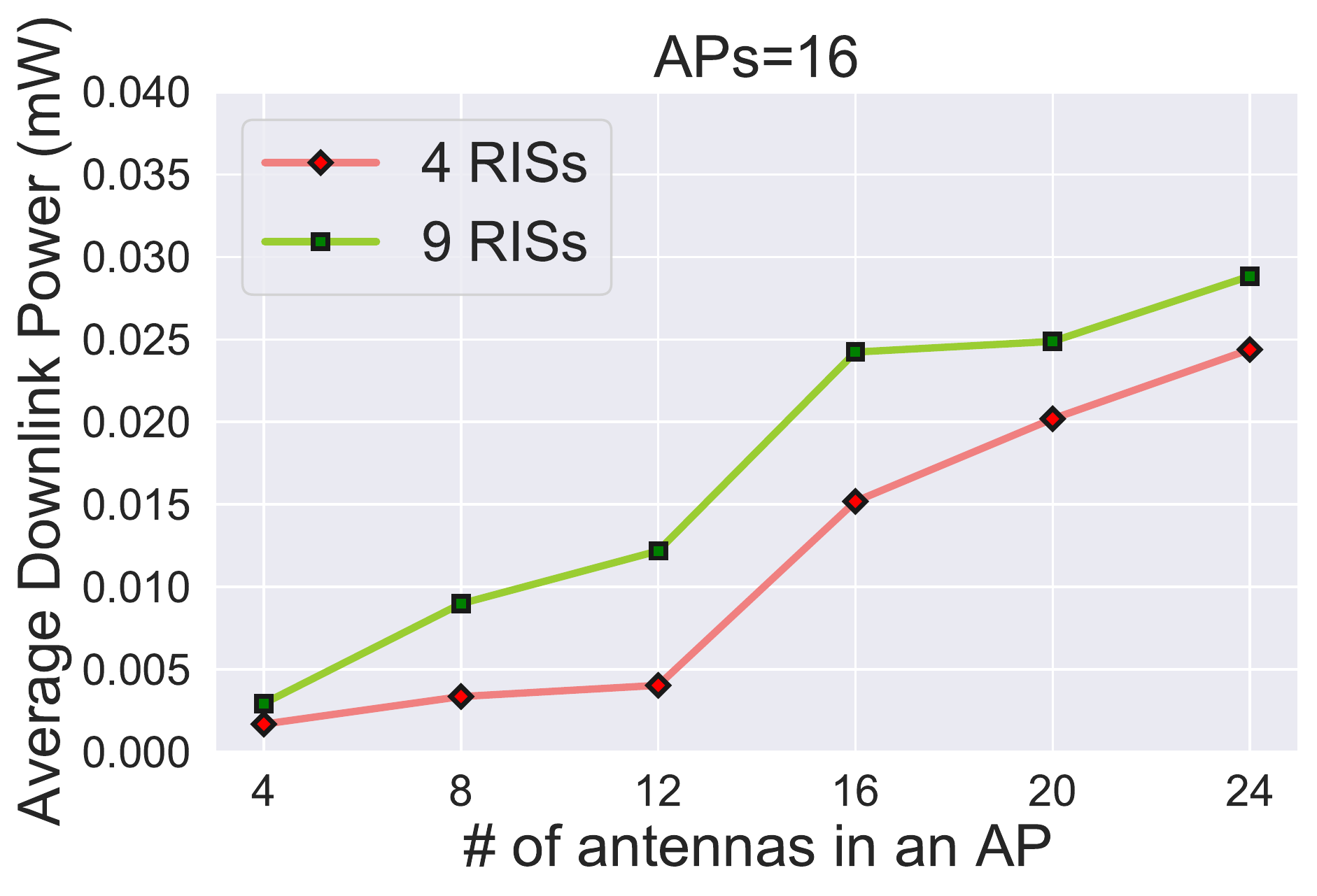}
        }
        \hspace{-5pt} 
        \subfigure[]
        {
        \label{fig:elements4}
            \includegraphics[width=0.23\textwidth, keepaspectratio=true]{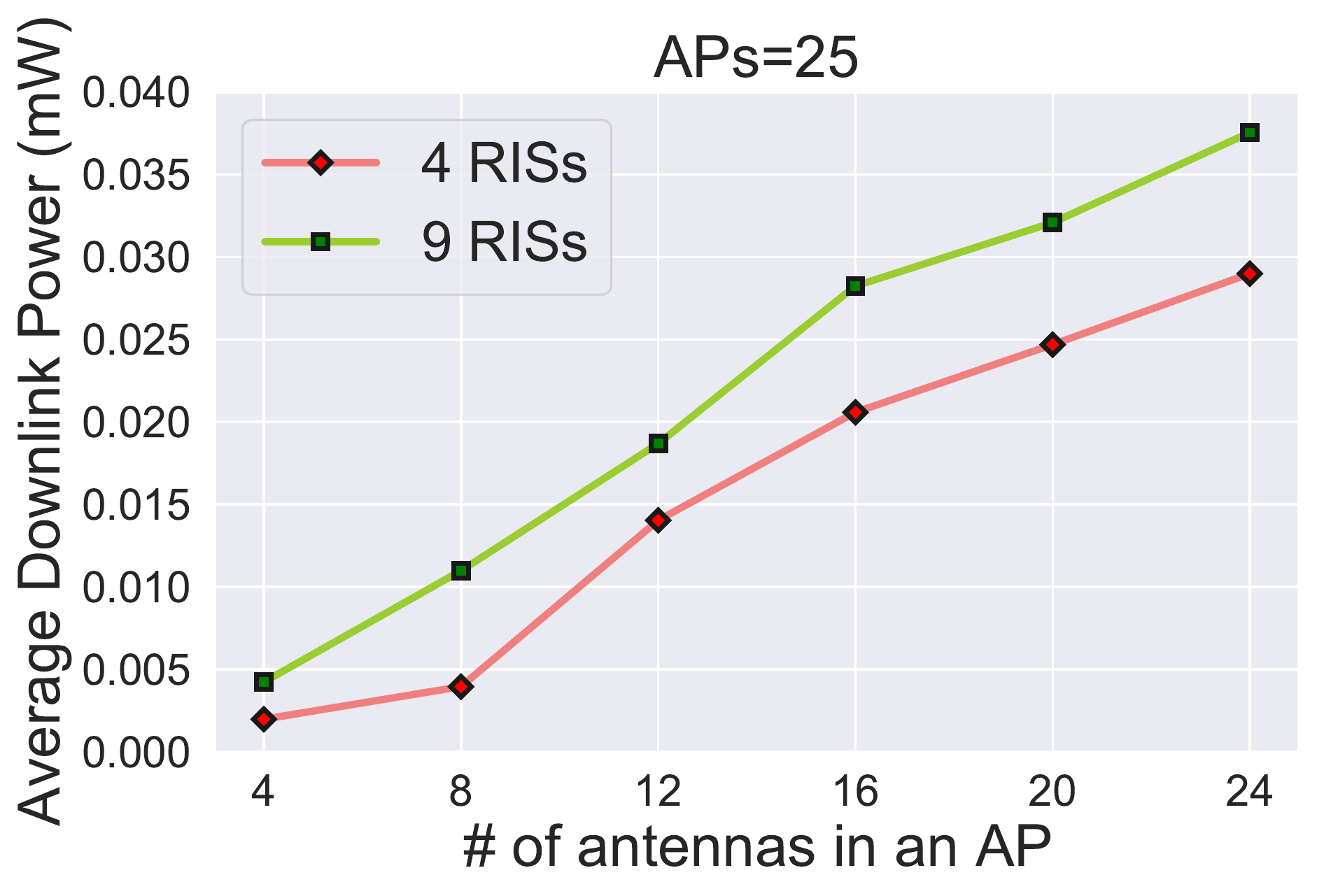}
        }
      
    \end{center}
    \vspace{-15pt}
    \caption{Harvested power with hybrid deployment strategy for (a-b) different number of reflecting elements in 9 and 16 RISs respectively, and (c-d) different number of antennas in 16 and 25 access points respectively. \done{The text in the figure needs to be larger.} \alvi{for this figure, I can have more graphs with varying number of RISs and APs. Also, we can use different system parameters.}} 
    \label{fig:elements}
    \vspace{-9pt}
\end{figure*}

\subsection{Implementation of the Energy Harvesting Model \done{The title is poor. What is this algorithm about?} \alvi{about the energy harvesting}}

Finally, in this section, we provide an elaborated description of the energy harvesting algorithm that we considered in this work, represented by Algorithm~\ref{algo}. First, we initialize the necessary system parameters ($Param_{sys}$) like realization, transmit power ($Pow_{transmit}$), block coherence ($Block_{coher}$) etc., according to the requirements of different experiments. Necessarily, we keep the values of some of the parameters fixed across all the experiments, which are provided in Table~\ref{tab:values}. We consider random placement for the client IoT devices in the coverage area, which are represented by each of the setups. We consider multiple setups so that we do not get biased performance results from a particular placement. For saving and later averaging the achieved rates of the energy harvesting model, we maintain a matrix of rates across the setups. For each setup, we calculate the channel gain and realization of the channel through the $SetupFunc$ function, where parameters of the APs ($Param_{AP}$) and RISs ($Param_{RIS}$) are passed as arguments. Then come up with the statistical terms and amount of harvested energy through the $ChannelEstimation$ function. Here we pass the system parameters, channel gain and realization of the channel as argument. Depending on the value of large-scale fading co-efficient ($\beta$), statistical terms and amount of harvested energy, the final rates are calculated by $SpectralEfficiency$ function. Then we check the feasibility of the rate solutions and terminate the iteration for a particular setup when feasible solutions are achieved. As mentioned earlier, we store the rates of different setups and calculate the average harvested energy for the model.

\begin{table}[!htb]
\centering
\caption{Values of different parameters for the \name{} framework.}
\label{tab:values}
\begin{tabular}{|l|c|}
\hline
\multicolumn{1}{|c|}{\textbf{Parameter Name}}                                                         & \textbf{Value Assigned}  \\ \hline
\textit{Pilot transmit power (W)}                                                                     & 10\textasciicircum{}(-7) \\ \hline
\textit{Total power limit per AP (W)}                                                                 & 10/U                     \\ \hline
\textit{Compute noise power (dBm)}                                                                    & -96                      \\ \hline
\textit{Length of coherence block}                                                                    & 200                      \\ \hline
\textit{Pilot Length}                                                                                 & 5                        \\ \hline
\textit{Number of downlink samples}                                                                   & 25                       \\ \hline
\textit{Carrier frequency (GHz)}                                                                      & 3.4                      \\ \hline
\textit{\begin{tabular}[c]{@{}l@{}}Standard deviation of \\ shadow fading for LOS (dB)\end{tabular}}  & 3                        \\ \hline
\textit{\begin{tabular}[c]{@{}l@{}}Standard deviation of \\ shadow fading for NLOS (dB)\end{tabular}} & 4                        \\ \hline
\end{tabular}
\end{table}


\section{Evaluation Results and Discussion}
\label{sec:evaluation}

\begin{figure*}[t]
\centering
\includegraphics[width=.95\textwidth]{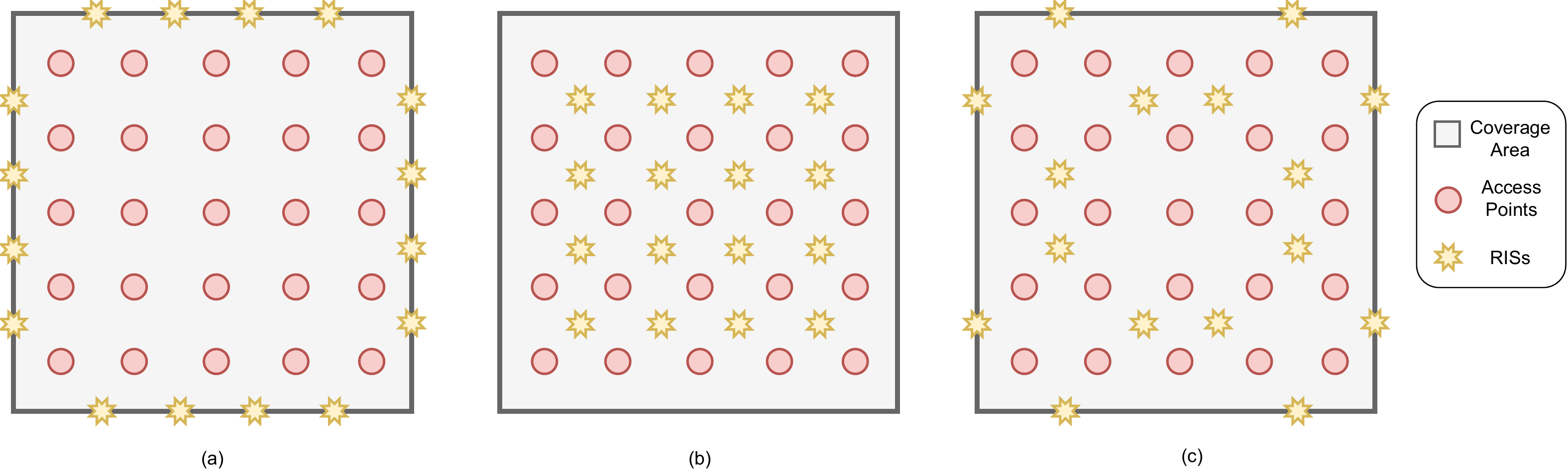}
\vspace{-12pt}
\caption{For a given area with 25 fixed access point locations, (a) edge deployment strategy, (b) central deployment strategy, and (c) hybrid deployment strategy, for 16 RISs. \alvi{I can not have more figures here}}
\label{fig:ds}
\vspace{-6pt}
\end{figure*}

\begin{figure*}[t]
    \begin{center}
    \hspace{-10pt} 
     \subfigure[]
        {
        \label{fig:harvested1}
            \includegraphics[width=0.32\textwidth, keepaspectratio=true]{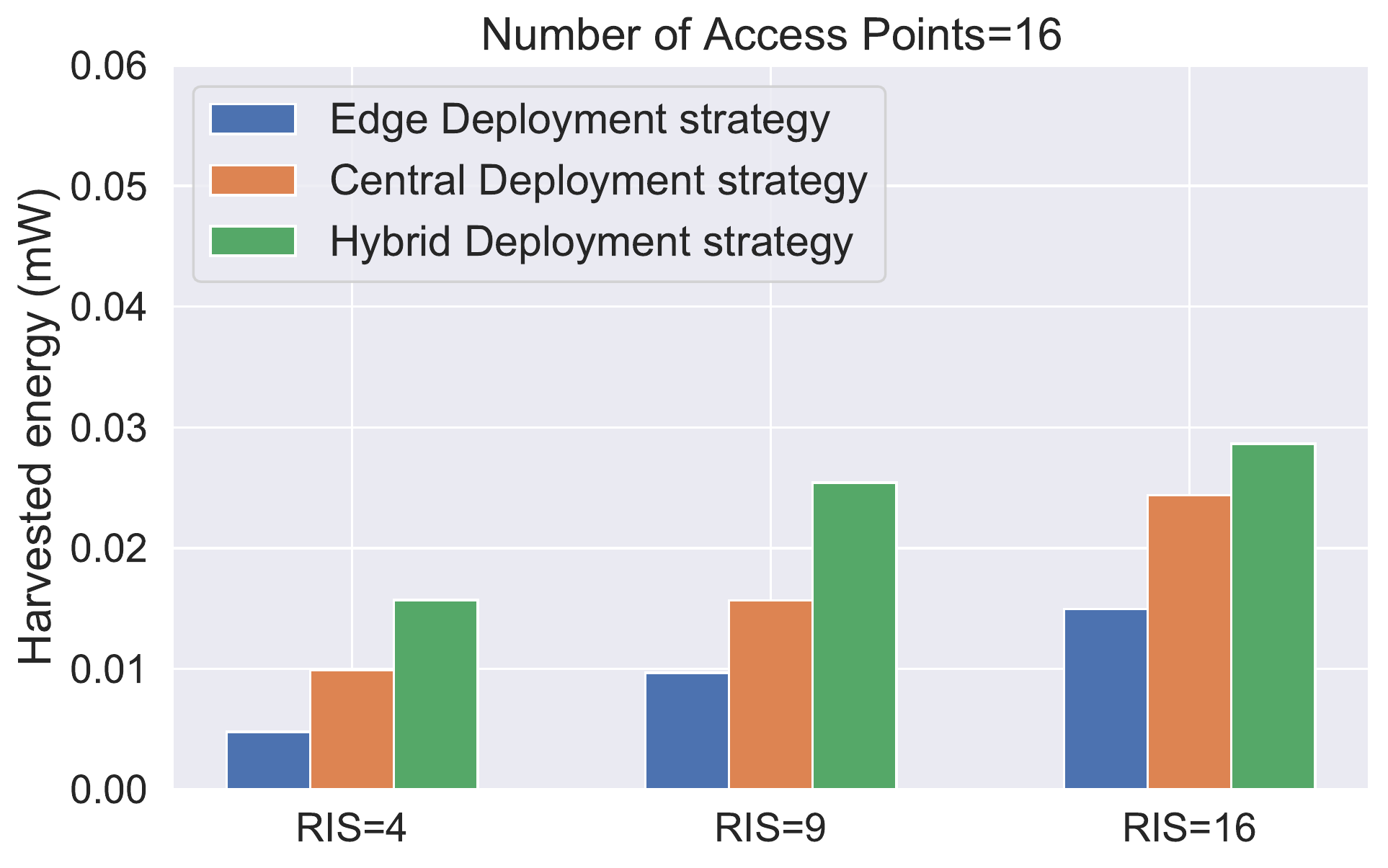}
        }     
        \hspace{-5pt}
    \subfigure[]
        {
        \label{fig:harvested2}
            \includegraphics[width=0.32\textwidth, keepaspectratio=true]{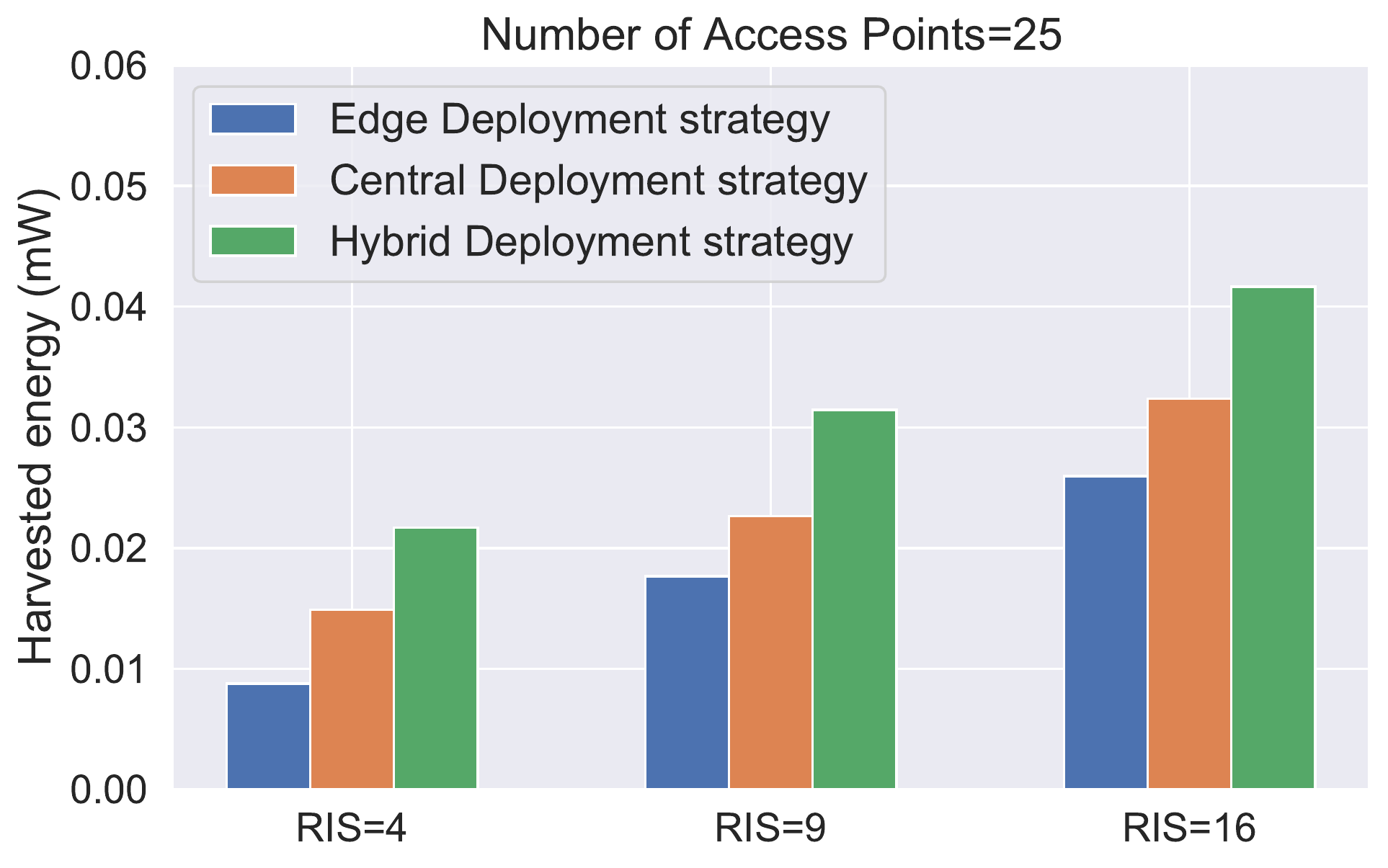}
        }     
        \hspace{-5pt}
    \subfigure[]
        {
        \label{fig:harvested3}
            \includegraphics[width=0.32\textwidth, keepaspectratio=true]{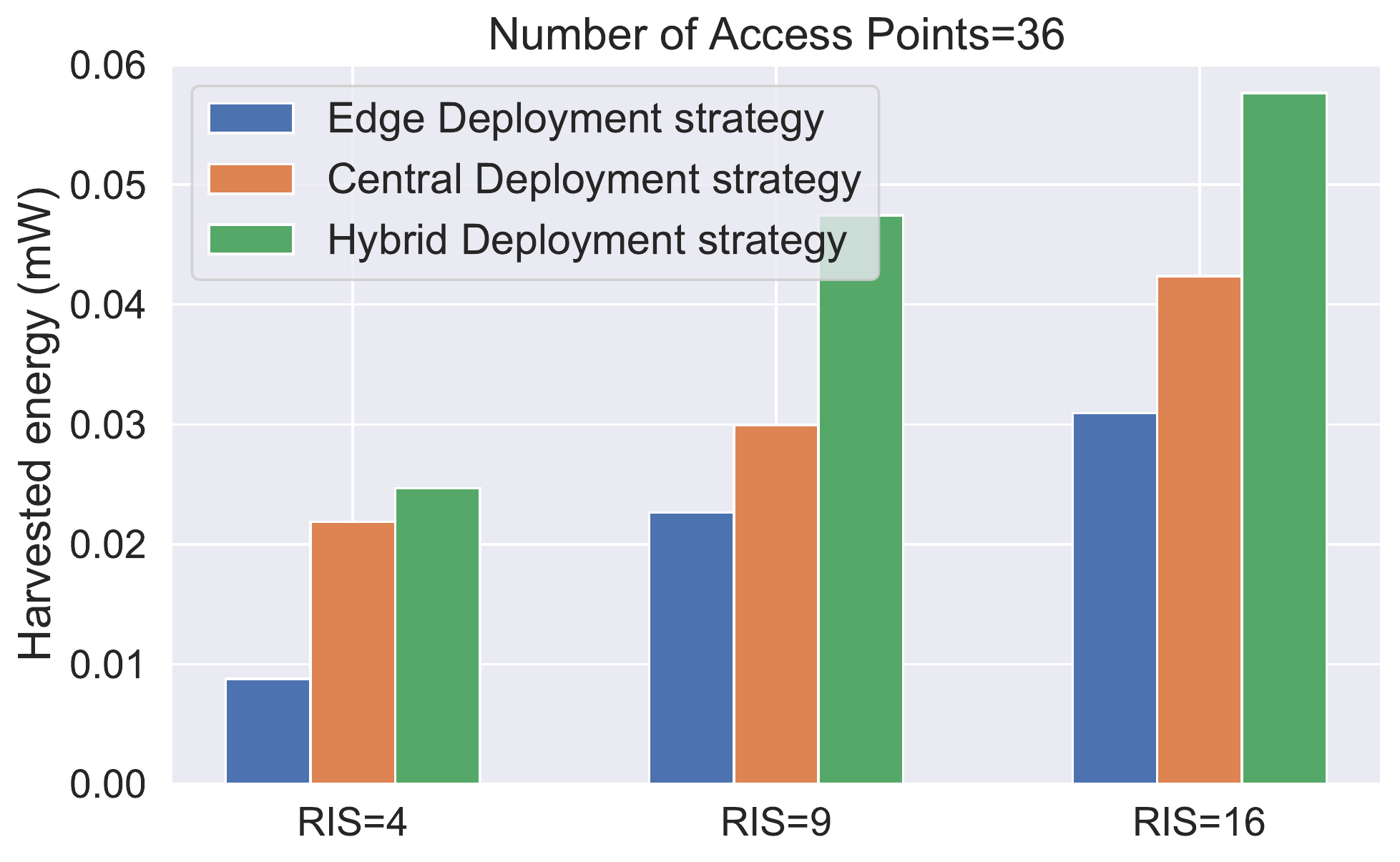}
        }        
      
    \end{center}
    \vspace{-15pt}
    \caption{Amount of harvested energy with different deployment strategies for (a) 16 access points, (b) 25 access points and (c) 36 access points, assisted by variable number of RISs. \alvi{I can have figures with fixed number of RISs and varying number of APs.}}
    \label{fig:harvested}
    \vspace{-9pt}
\end{figure*}

\rahman{Can you think for extending the evaluation results -- more graphs?} \alvi{Sure, I can play around the different system parameters to add more graphs.}

In this section, we present the experimental setup and necessary evaluation metrics to assess the energy harvesting performance of the proposed \name{} framework. We leverage the implementation of MMF optimization from \cite{demir2020joint}, to calculate the downlink RF energy harvesting from the CFmMIMO UAVs. We further extend it by introducing RIS-assisted energy harvesting. We validate the claim of achieving an increased amount of harvested energy with RIS assistance through comparison with regular MMF and RIS assisted MMF, with respect to empirical cumulative distribution function value of the harvested energy. Later, we experiment with a variable number of meta elements in each RIS and try out with different number of access points to assist in order to find out the optimal number of meta elements for a given number of access points. Also, we try with different number of antennas in an AP with a variable number of RISs and try to find out the optimal number of antennas for a given number of RISs. Finally, we experiment with three different deployment strategies for the RISs and try with different numbers of APs to come up with the best deployment strategy.

\subsection{Spectral Efficiency per user}
In this section, we evaluate the performance of \name{} framework by comparing the empirical cumulative distribution function of individual spectral efficiency per user, achieved by the regular MMF mechanism and RIS-assisted MMF mechanism, as represented in Fig.~\ref{fig:cdf_mmf}. 
We compare the spectral efficiency for all the combinations of coherent, non-coherent and linear, non-linear energy transmission. 
For the coherent linear energy transmission in Fig.~\ref{fig:cdf_mmf1}, it can be seen that RIS-assisted MMF is achieving a higher amount of spectral efficiency compared to standard MMF. Similarly, from Fig.~\ref{fig:cdf_mmf2}, Fig.~\ref{fig:cdf_mmf3}, Fig.~\ref{fig:cdf_mmf4}, it is observed that RIS assisted MMF achieved higher spectral efficiency per user for other types of signal as well. Both the models show similar distributions; for coherent linear signal, the distribution is more skewed than the non-coherent non-linear signal. The upper tail of the curves represents the users with good channel conditions, who achieve higher spectral efficiency than the other users because of their operation in the saturation region. Fig.~\ref{fig:nap} represents the comparison of performance between the MMF and proposed \name{} framework with respect to the harvested energy. It is evident from Fig.~\ref{fig:nap1} and Fig.~\ref{fig:nap3} that \name{} achieves substantially more harvested energy than MMF for both coherent linear and non-coherent non-linear energy transmissions. To get a concrete idea about the improvement, we increased the number of RISs leveraged for both types of signal in Fig.~\ref{fig:nap2} and Fig.~\ref{fig:nap4}. 
It is observed that the improvement is exponential with the increased number of RISs.

\subsection{Optimal number of meta-element and antennas}
In this part, we experiment with different numbers of meta elements in the RISs, assisting variable number of APs and different number of antennas in the APs assisted by a variable number of RISs to come up with the optimal choices, as shown in Fig.~\ref{fig:elements}. We keep the number of RISs and APs fixed for the variable meta elements and variable antennas case, respectively. From Fig.~\ref{fig:elements1}, it is observed that with 9 access points, the amount of harvested energy is very low, even with a higher number of meta elements in each of the 9 RISs. On the other hand, with 16 access points, the amount harvested energy seems to have an upward trend with the increasing number of meta elements in the RISs. On the other hand, in Fig.~\ref{fig:elements2}, a more steep curve is observed with 16 access points as the number of RISs is increased to 16. In Fig.~\ref{fig:elements3} and Fig.~\ref{fig:elements4} it is seen that varying the number of RISs doesn't have that much impact compared to a varying number of APs. In both the figures, for 4 RISs, the incremental number of antennas incurred a higher amount of harvested power. For 9 RISs, the trend is similar, but the magnitude of the downlink power is lower. 

\subsection{Best deployment strategy}
We experiment with three different deployment strategies for the RISs and try with different numbers of APs to come up with the best deployment strategy. The strategies include: edge deployment, central deployment, and hybrid deployment, as shown in Fig.~\ref{fig:ds}. In the edge deployment (Fig.~\ref{fig:ds}(a)), all the RISs are placed at the edges of the coverage area. In the second deployment strategy, the RISs are placed in between the access points, as presented in Fig.~\ref{fig:ds}(b). In the final deployment strategy, the RISs are deployed in a mixed manner, where half of them are placed at the edge and the other half are placed in the middle of the access points (Fig.~\ref{fig:ds}(c)). From Fig.~\ref{fig:harvested}, it can be observed that the hybrid deployment is optimal strategy for all different choices of RIS with 16 APs (Fig.~\ref{fig:harvested1}), 25 APs (Fig.~\ref{fig:harvested2}) and 36 APs (Fig.~\ref{fig:harvested3}). One interesting trend can be found from the figures, that is, with the increase of APs in the coverage areas, the hybrid deployment strategy is becoming exponentially prominent than the other strategies.


\section{Conclusion}
\label{sec:conclusion}

In this work, we have presented a comprehensive RF energy harvesting framework that leverages the combined benefits of UAV mounted CFmMIMO and RIS. The tethered UAVs, operating in a cell-free fashion, provide strong LoS signal in the general coverage area, while untethered ones provide service in peak areas for their flexible deployment. We have devised a mechanism to utilize RISs for directive signals towards the target devices, which will assist both energy harvesting and information transfer. The empirical evaluation results have validated that our framework can achieve an increased amount of energy harvesting can be achieved than the MMF~\cite{demir2020joint}. We have also studied different deployment strategies for the RISs, and after comparing their energy harvesting performance, we have found the best strategy properties. In our future work, we will try to optimize the deployment positions and heights of both the UAV APs and RISs for even improved energy harvesting than the best deterministic strategy.


\bibliographystyle{IEEEtran}
\bibliography{References}

\end{document}